\documentclass[fleqn,usenatbib]{mnras}

\usepackage{newtxtext,newtxmath}

\usepackage[T1]{fontenc}

\DeclareRobustCommand{\VAN}[3]{#2}
\let\VANthebibliography\thebibliography
\def\thebibliography{\DeclareRobustCommand{\VAN}[3]{##3}\VANthebibliography}

\usepackage{graphicx}	
\usepackage{amsmath}	
\usepackage{comment}
\usepackage{booktabs}  
\providecommand{\abs}[1]{\lvert#1\rvert}

\usepackage{caption}
\usepackage[flushleft]{threeparttable} 
\usepackage{xspace}

\newcommand{\msun}{M$_{\sun}$\xspace}
\newcommand{\rsun}{R$_{\sun}$}

\newcommand{\h}{^{\rm h}}
\newcommand{\m}{^{\rm m}}

\newcommand{\Pobs}{$P_{\rm obs\ }$}

\newcommand{\Pdotobs}{$\dot{P}_{\rm obs\ }$}
\newcommand{\Pdotint}{$\dot{P}_{\rm int}$}
\newcommand{\PRESTO}{\texttt{PRESTO}}
\newcommand{\PULSARMINER}{\texttt{PULSAR\_MINER}}
\newcommand{\TEMPOTWO}{\texttt{TEMPO2}}

\newcommand{\DRACULA}{\texttt{DRACULA}}
\newcommand{\SeeKAT}{\texttt{SeeKAT}}

\title[Pulsars in M62]{Discoveries and Timing of Pulsars in M62}

\author[L. Vleeschower et al.]{\parbox{\textwidth}{
L.~Vleeschower,$^{1,2}$\thanks{E-mail: laila.vleeschowercalas@postgrad.manchester.ac.uk}A.~Corongiu,$^{3}$
B.~W.~Stappers,$^{1}$
P.~C.~C.~Freire,$^{4}$
A. Ridolfi,$^{3,4}$
F.~Abbate,$^{3,4}$
S.~M.~Ransom,$^{5}$
A.~Possenti,$^{3}$
P.~V.~Padmanabh,$^{6,7,4}$
V.~Balakrishnan,$^{4}$
M.~Kramer,$^{4,1}$
V.~Venkatraman~Krishnan,$^{4}$
L.~Zhang,$^{8,9}$
M.~Bailes,$^{9,10}$
E.~D.~Barr,$^{4}$
S.~Buchner,$^{11}$
and W.~Chen$^{4}$
}
\\ \\ \\ \\
\parbox{\textwidth}{
$^{1}$Jodrell Bank Centre for Astrophysics, Department of Physics and Astronomy, The University of Manchester, Manchester M13 9PL, UK\\
$^{2}$Center for Gravitation, Cosmology, and Astrophysics, Department of Physics, University of Wisconsin-Milwaukee, P.O. Box 413, Milwaukee, WI 53201, USA\\
$^{3}$ INAF-Osservatorio Astronomico di Cagliari, Via della Scienza 5, 09044, Selargius, Italy\\
$^{4}$Max-Planck-Institut f\"{u}r Radioastronomie, Auf dem H\"{u}gel 69, D-53121 Bonn, Germany\\
$^{5}$NRAO, 520 Edgemont Rd., Charlottesville, VA, 22903, USA\\
$^{6}$Max-Planck-Institut f\"{u}r Gravitationsphysik (Albert-Einstein-Institut), D-30167 Hannover, Germany\\
$^{7}$Leibniz Universit\"{a}t Hannover, D-30167 Hannover, Germany\\
$^{8}$National Astronomical Observatories, Chinese Academy of Sciences, A20 Datun Road, Chaoyang District, Beijing 100101, People's Republic of China\\
$^{9}$Centre for Astrophysics and Supercomputing, Swinburne University of Technology, P.O. Box 218, Hawthorn, VIC 3122, Australia\\
$^{10}$ ARC Centre of Excellence for Gravitational Wave Discovery (OzGrav)\\
$^{11}$ South African Radio Astronomy Observatory, 2 Fir Street, Black River Park, Observatory 7925, South Africa \\
}
}

\date{Accepted XXX. Received YYY; in original form ZZZ}

\pubyear{2023}

\begin{document}
\label{firstpage}
\pagerange{\pageref{firstpage}--\pageref{lastpage}}
\maketitle

\begin{abstract}
Using MeerKAT, we have discovered three new millisecond pulsars (MSPs) in the bulge globular cluster M62: M62H, M62I, and M62J. All three are in binary systems, which means all ten known pulsars in the cluster are in binaries. M62H has a planetary-mass companion with a median mass $M_{\rm c,med} \sim 3$\,M$_{\rm J}$ and a mean density of $\rho \sim 11$\,g\,cm$^{-3}$. M62I has an orbital period of 0.51\,days and a $M_{\rm c,med} \sim 0.15$\,\msun. Neither of these low-mass systems exhibit eclipses. M62J has only been detected in the two UHF band (816\,MHz) observations with a flux density $S_{816} = 0.08$\,mJy. The non-detection in the L-band (1284\,MHz) indicates it has a relatively steep spectrum ($\beta < -3.1$). We also present 23-yr-long timing solutions obtained using data from the Parkes ``Murriyang'', Effelsberg and MeerKAT telescopes for the six previously known pulsars. For all these pulsars, we measured the second spin-period derivatives and the rate of change of orbital period caused by the gravitational field of the cluster, and their proper motions. From these measurements, we conclude that the pulsars’ maximum accelerations are consistent with the maximum cluster acceleration assuming a core-collapsed mass distribution. Studies of the eclipses of the redback M62B and the black widow M62E at four and two different frequency bands, respectively, reveal a frequency dependence with longer and asymmetric eclipses at lower frequencies. The presence of only binary MSPs in this cluster challenges models which suggest that the MSP population of core-collapsed clusters should be dominated by isolated MSPs.  


\end{abstract}


\section{Introduction}
\label{s:Intro}

Globular Clusters (GCs) are considered prime locations for discovering millisecond pulsars (MSPs) due to their ultra-dense cores, which promote the formation of binary systems containing neutron stars (NSs). Within these systems, the neutron star can undergo spin-up through mass accretion from its evolving companion. MSPs that arise from these scenarios are highly valuable for investigating the dynamical interactions in GC cores, studying the evolution of embedded binary systems, and examining the properties of the intra-cluster medium \citep[e.g.][]{Freire+2001,Abbate+2018,Abbate+2019}, but also for giving constraints on the NSs masses \citep[e.g.][]{Freire+2008a, Freire+2008b} and enable tests of gravity theories \citep{Jacoby+2006}. Nevertheless, finding MSPs in clusters is a challenging endeavour, primarily due to many being at large distances, resulting in weak pulsed emission and distorted signals caused by propagation through the dispersive interstellar medium. Additionally, the apparent rapid changes in period due to binary motion and periodic eclipses of the radio signal further complicate the search process.

M62 (also known as NGC~6266) is a metal-poor globular cluster with a metallicity of [Fe/H] = -1.075 $\pm$ 0.039 \citep{Lapenna+2015}. It is located at right ascension $\alpha = 17\h01\m12\fs9$ and declination $\delta = -30\degr06\arcmin48\farcs2$, placing it towards the Galactic bulge, in the southern region of the equatorial constellation Ophiuchus, at galactic coordinates $l = 353.575^{\circ}$ and $b = 7.317^{\circ}$. It is at a distance of 6.03(9)\,kpc\footnote{Distance reported in \url{https://people.smp.uq.edu.au/HolgerBaumgardt/globular/parameter.html} (accessed on 28 August 2023) obtained by fitting the cluster with a grid of \textit{N}-body models as in \citet{Baumgardt+Vasiliev2021}, driven by the new cluster velocities from \citet{Martens+2023}, Baumgardt, H. private communication. \\ All positional and size parameters of the cluster used in this work were obtained from the same website. We will refer to this website as Baumgardt's catalogue hereafter.} from the Sun. M62 is recognised for having one of the highest stellar encounter rates among globular clusters \citep{Pooley+2003}. It is also among the 10\% of the most massive and dense globular clusters, with a total mass of 5.81(4) $\times 10^5$\,\msun, and possesses a core radius of $r_{\rm c} = 0.23$\,pc ($0.13$\,arcmin) and a half-mass radius of $r_{\rm h} = 2.91$ pc ($1.66$\,arcmin).

This cluster was previously classified as being core-collapsed \citep{Djorgovski+Meylan1993}, however more recent studies had ruled out this hypothesis \citep{Beccari+2006}. This was supported by the fact that all pulsars in the cluster are in circular binary systems, which is not expected for core-collapsed GCs \citep{Verbunt+Freire2014}. However, the latest results of the structural parameters of this cluster from Baumgardt's catalogue, indicate a smaller core radius than previously stated, resulting in a high value of its central concentration ($c=2.35$). In the context of the classification proposed by \citealt{Meylan+Heggie1997}, this suggests that the cluster may be considered as either collapsed or on the verge of collapsing.


Forty-three X-ray sources, including a black hole candidate \citep{Chomiuk+2013}, have been detected within the half-light radius of the cluster using Chandra \citep{Oh+2020}. Additionally, M62 is known to host seven pulsars, all of which are in binary systems and with spin periods smaller than 8\,ms. The first pulsar discovered in this cluster was M62A, a binary system with a spin period of 5.24\,ms and an orbital period of 3.8\,days \citep{DAmico+2001}. Preliminary analysis in \citet{DAmico+2001b} revealed the discovery of two more binary pulsars, M62B and M62C, characterised by short orbital periods of 3.8 and 5.2\,hr, respectively. These three pulsars were discovered using the Parkes ``Murriyang'' telescope. Subsequently, three additional pulsars, namely M62D-F, were discovered using the 100-m Green Bank Telescope \citep[GBT;][]{Jacoby+2002}. M62D has a spin period of 3.42\,ms, an orbital period of 1.1\,days, and a minimum companion mass of 0.12\,\msun. M62E is a 3.23-ms binary pulsar in a 3.8\,hr circular orbit, with a companion mass of at least 0.03\,\msun. M62F has a spin period of 2.29\,ms, an orbital period of 4.8\,hr, and a minimum companion mass of 0.02\,\msun. Notably, pulsars M62B and M62E exhibit eclipses, with M62B classified as a redback and M62E as a black widow. The mass of its companion suggests M62F is another black widow pulsar, although eclipses have not been observed for this particular pulsar.

The most recent discovery in the cluster was made with observations from the South African MeerKAT radio telescope \citep{Booth+Jonas2012} using only the 44 antennas located in the inner 1-km core of the array \citep{Ridolfi+2021}. Pulsar M62G has a spin period of 4.61\,ms and is situated within a binary system characterised by an orbital period
 of 0.8\,days, and a companion with a minimum mass of 0.1\,\msun. 
 
In this paper, we present the discovery of three more pulsars in this cluster using the high gain/low system temperature MeerKAT radio telescope and the updated and long-term timing solutions of all the previously known pulsars making use of archival Parkes ``Murriyang'', Effelsberg and MeerKAT data. It is structured as follows: the observations and data reduction are described in Section \ref{s:Obs and Data}. We report the discoveries in Section \ref{s:Discoveries}. The timing solutions of all the previously known and the new pulsars are presented in Section \ref{s:Timing}. In Section \ref{s:Discussion} the results and their implications are summarised. Finally, the conclusions are presented in Section \ref{s:Conclusions}.



\section{Observations and Data Reduction}
\label{s:Obs and Data}

The observations used in the analysis of this work have been taken with the Parkes ``Murriyang'', Effelsberg and MeerKAT radio telescopes covering a total time span of over 23\,yr. The data, observing systems and data reduction are described below.

\subsection{Parkes and Effelsberg observations and data analysis}

Observations of M62 taken with the Parkes ``Murriyang'' radio telescope, were carried out during an epoch range of about one decade, namely from 2000 June to 2011 July. These observations were taken with the Multibeam and the H$-$OH receivers, depending on availability, at the central frequency of 1369\,MHz with a bandwidth of 256\,MHz. The two detected orthogonal polarisations were summed in quadrature and 1-bit digitised by using the Analog Filterbank (AFB; see e.g. \citealt{DAmico+2001}).

We also include in our analysis five observations taken from 2015 June to 2015 August 16 with the Effelsberg telescope (project id 16-15). These observations were made with the Ultra-Broad-Band (UBB) receiver, which allowed the simultaneous observation of three frequency bands (601--695 MHz, 836--930 MHz and 1148--1460 MHz) with the purpose of characterising the eclipses of pulsar M62B (see Section \ref{s:Timing}).

\subsection{MeerKAT observations and data analysis}

M62 was previously studied in the context of the MeerKAT GC census through the MeerTime project \citep[see][]{Ridolfi+2021}. The cluster was subsequently targeted using TRAPUM\footnote{http://www.trapum.org} (TRAnsients and PUlsars with MeerKAT; \citealt{Stappers+Kramer2016}), allowing us to generate hundreds of coherent beams and enabling us to cover a larger area of the cluster using the full MeerKAT sensitivity. We conducted  seven observations under the TRAPUM project, between 2020 October and 2022 December, employing at least 56 antennas. The full list of MeerKAT observations used for this work is reported in Table \ref{t:obs_summary}. 

We employed the Filterbanking Beamformer User Supplied Equipment \citep[FBFUSE;][]{Barr2018}, which facilitated the formation of 275-288 synthesised beams, covering a region of up to $\sim 3$\,arcmin in radius centred on the nominal cluster centre. The shape and tiling pattern of these beams were estimated using the \texttt{Mosaic} software \citep{Chen+2021}. The exact size, shape and orientation of the beams depend on several factors, such as the number of antennas used, the elevation, and the observing time and frequency. The data were recorded as filterbanks with total intensity derived from the two orthogonal polarisations. Five of the observations were done using the L-band (856--1712\,MHz) receivers, centred at a frequency of 1284\,MHz, employing at least 56 antennas with a $> 70$\% overlap\footnote{Determines the sensitivity level at which neighbouring beams intersect.} of the synthesised beams \citep{Chen+2021}. The raw data set consisted of 4096 frequency channels across a bandwidth of 856\,MHz sampled every 76\,$\mu$s. Two more TRAPUM observations were recorded with the Ultra High Frequency (UHF; 544-1088\,MHz) receivers, employing 56 antennas and the beams placed with an 85\% overlap. These observations were centred at a frequency of 816\,MHz with total intensity formed from the two orthogonal polarisations. The data set also comprised 4096 frequency channels but with a bandwidth of 544\,MHz and sampled every 60\,$\mu$s. Figure \ref{fig:M62_psrpositions} shows an example of the beams at both L-band and UHF observations using 56 antennas at their 50 per cent power level.

The MeerTime observations piggybacked on some of the TRAPUM observations (see Table \ref{t:obs_summary}). This was with the purpose of getting polarisation information and for future searches in the single MeerTime beam. These will be presented elsewhere. However, we used them to get a precise measurement of the time offset between the two different backends (see Section \ref{s:Timing}). 

Furthermore, an orbital campaign was carried out with the purpose of determining the orbits of the newly discovered pulsars and also to phase-connect the solution of pulsar M62G. This pulsar was believed to have a partial phase connection \citep{Ridolfi+2021} but after adding the subsequent observations (including all those before the campaign, see Table \ref{t:obs_summary}) the connection was not clear. The campaign consisted of five L-band observations of two hours each, over two consecutive days (in order to optimally sample the orbits based on the initial estimates for the new pulsars) as part of the MeerTime project \citep{Bailes+2020}. Four PTUSE (Pulsar Timing User Supplied Equipment)/B-engine beams were used and pointed at the four pulsars of interest (M62G-J). These observations were piggybacked with TRAPUM, pointing one FBFUSE beam to each of the known pulsars and M62H (which had a \SeeKAT~localisation, see Section \ref{s:Discovery_M62H}), and seven beams were pointed towards the poorly localised (at that time) M62G, M62I and M62J. We only discuss the use of TRAPUM data from this campaign here.

Recording and pre-processing of the beams was carried out using the APSUSE (Acceleration Pulsar Search User Supplied Equipment) computing cluster. Once the recording phase was completed, the data underwent incoherent de-dispersion using a DM of 114.00\,pc\,cm$^{-3}$, which corresponds to the median DM of the then-known pulsars within the cluster. The data were later cleaned of bright Radio Frequency Interference (RFI) using the Inter-Quartile Range Mitigation algorithm\footnote{\url{https://github.com/v-morello/iqrm}} \citep{Morello+2022}; later observations also included the use of \texttt{filtool} from \texttt{PulsarX}\footnote{https://github.com/ypmen/PulsarX} \citep{Men+2023} to remove RFI. After de-dipsersion, the number of frequency channels was reduced to 256 for the following analysis and long-term storage.

\begin{table*}
\caption{Summary of the MeerKAT observations used in this analysis. The Date/Times are specified in UTC. Columns 10 and 11 display the observations in which the eclipse is seen for pulsars M62B and M62E, respectively, categorised as follows: `B' refers to observations in which both the ingress and egress are visible, `I' to ingress only, and `E' to egress only.}
\label{t:obs_summary}
\centering
\resizebox{\textwidth}{!}{
\begin{tabular}{ccccccccccc}
\hline \hline 
\multicolumn{1}{c}{Date/Time}   & \multicolumn{1}{c}{Start MJD} & \multicolumn{1}{c}{f$_{\rm c}$}  & \multicolumn{1}{c}{BW} & \multicolumn{1}{c}{T$_{\rm obs}$} & \multicolumn{1}{c}{$N_{\rm{ant}}$} & \multicolumn{1}{c}{H Det }  & \multicolumn{1}{c}{I Det } & \multicolumn{1}{c}{J Det} & M62B & M62E \\ 
(yy-MM-dd-HH:mm) & & (MHz) & (MHz) & (min) & & Y/N & Y/N & Y/N & \\
\hline 
19-04-29-19:40$^m$ & 58602.82 & 1122 & 321 & 75.6 & 57  & N & Y & N & B & -- \\
19-04-29-19:40$^m$ & 58602.82 & 1444 & 321 & 75.6 & 57 & N & Y & N & B & -- \\
19-10-13-10:31$^m$ & 58769.44 & 1284 & 642 & 150 & 52 & Y & Y & N & B & B \\
19-11-15-10:49$^m$ & 58802.45 & 1284 & 642 & 150 & 64  & N & Y & N & B & E \\
19-12-01-10:13$^m$ & 58818.43 & 1284 & 642 & 150 & 61 & Y & Y & N & B & B \\
19-12-27-07:57$^m$ & 58844.33 & 1284 & 642 & 24.2 & 41 & N & Y & N & -- & -- \\
19-12-27-08:28$^m$ & 58844.35 & 1284 & 642 & 130 & 41 & Y & Y & N & B & -- \\
20-02-16-00:25$^m$ & 58895.02 & 1284 & 856 & 130 & 52 & Y & Y & N & -- & -- \\
20-02-21-23:54$^m$ & 58900.99 & 1284 & 856 & 210 & 52 & Y & Y & N & B & B \\
20-02-22-23:57$^m$ & 58901.99 & 1284 & 856 & 80 & 51 & Y & Y & N & -- & -- \\
20-02-23-01:20$^m$ & 58902.05 & 1284 & 856 & 123 & 50 & Y & Y & N & B \\
20-02-25-00:04$^{m*}$ & 58904.00 & 1284 & 856 & 182 & 50 & Y & Y & N & B & B \\
20-08-13-18:52$^{m*}$ & 59074.78 & 1284 & 856 & 120 & 40 & Y & Y & N & B & I \\
20-10-14-14:58$^m$ & 59136.62 & 1284 & 856 & 240 & 38 & Y & Y & N & B & B \\
20-10-14-15:03$^t$ & 59136.62 & 1284 & 856 & 234.8 & 56 & \,\,\,Y$^s$ & Y & N & B & B \\
20-10-23-14:02$^m$ & 59145.58 & 1284 & 856 & 240 & 41 & N & Y & N & B & E \\
20-11-01-11:03$^m$ & 59154.46 & 1284 & 856 & 240 & 41 & Y & Y & N & B & B \\
20-11-01-11:03$^t$ & 59154.46 & 1284 & 856 & 239.5 & 56 & \,\,\,Y$^s$ & \,\,\,Y$^s$ & N & B & B \\
21-06-12-19:58$^m$ & 59377.83 & 1284 & 856 & 115 & 63 & N & Y & N & I & B \\
21-09-21-15:55$^m$ & 59478.66 & 1284 & 856 & 120 & 38 & Y & Y & N & E & E \\
21-09-21-15:55$^t$ & 59478.66 & 1284 & 856 & 119.5 & 56 & \,\,\,Y$^s$ & \,\,\,Y$^s$ & N & E & E \\
21-12-30-11:59$^m$ & 59578.49 & 1284 & 856 & 120 & 44 & Y & Y & N & B & -- \\
21-12-30-12:27$^t$ & 59578.51 & 1284 & 856 & 92 & 60 & \,\,\,Y$^s$ & \,\,\,Y$^s$ & N & B & -- \\
22-04-20-23:13$^t$ & 59689.96 & 816 & 544 & 60 & 56 & \,\,\,Y$^s$ & \,\,\,Y$^s$ & \,\,\,Y$^s$ & B & -- \\
22-04-21-01:55$^t$ & 59690.08 & 1284 & 856 & 60 & 60 & \,\,\,Y$^s$ & \,\,\,Y$^s$ & N & B & E \\
22-12-01-14:20$^t$ & 59914.59  & 816 & 544 & 120 & 56 & \,\,\,Y$^s$ & \,\,\,Y$^s$ & \,\,\,Y$^s$ & B & B \\
23-07-04-14:57$^c$ & 60129.62 & 1284 & 856 & 120 & 60 & \,\,\,Y$^s$ & \,\,\,Y$^s$ & N & B & B \\
23-07-04-19:46$^c$ & 60129.82 & 1284 & 856 & 120 & 60 & \,\,\,Y$^s$ & \,\,\,Y$^s$ & N & B & B \\
23-07-05-12:22$^c$ & 60130.02 & 1284 & 856 & 120 & 60 & \,\,\,Y$^s$ & \,\,\,Y$^s$ & N & -- & -- \\
23-07-05-16:57$^c$ & 60130.70 & 1284 & 856 & 120 & 56 & \,\,\,Y$^s$ & \,\,\,Y$^s$ & N & B & -- \\
23-07-05-22:56$^{c}$ & 60130.96 & 1284 & 856 & 120 & 56 & \,\,\,Y$^s$ & \,\,\,Y$^s$ & N & B & E \\
\hline \hline 
\multicolumn{8}{l}{$^*$ These observations were mistakenly reported to have a duration of 210\,min in \citealt{Ridolfi+2021}.}\\
\multicolumn{8}{l}{$^c$ MeerTime/TRAPUM orbital campaign observations.}\\
\multicolumn{8}{l}{$^m$ MeerTime observations.}\\
\multicolumn{8}{l}{$^s$ pulsar found in the search.}\\
\multicolumn{8}{l}{$^t$ TRAPUM observations.} 
\end{tabular}
}
\end{table*}

All the data from the seven TRAPUM observations were analysed following a typical acceleration search method using the \PULSARMINER\ pipeline\footnote{\url{https://github.com/alex88ridolfi/PULSAR_MINER}}, (v.1.1.5, see further details in \citealt{Ridolfi+2021}) based on \PRESTO\footnote{PulsaR Exploration and Search TOolkit, https://github.com/scottransom/presto}~\citep{Ransom2001}. For the L-band observations, we generated 245 de-dispersed time series from a DM of 108.00\,pc\,cm$^{-3}$ to a DM of 120.20 in steps of 0.05\,pc\,cm$^{-3}$. The DM range was determined by considering a deviation of $\pm 5$\% of the mean DM of the then-known pulsars. The chosen DM step size ensured minimal DM smearing caused by an incorrect DM and sensible requirements for the processing. Regarding the UHF observations, we generated 1201 de-dispersed time series using the same DM range as for the L-band data, in steps of 0.01\,pc\,cm$^{-3}$.

The search encompassed spin periods ranging from 1\,ms to 20\,s for all the 288 beams for the first two TRAPUM observations, and only the central beams (7-beam tiling around the centre of the GC) of the subsequent observations, giving a total of 611 beams across all observations. All beams were analysed for both isolated and binary pulsars in the Fourier domain using a maximum number $z_{\max}$ of adjacent Fourier bins $z_{\rm max}=0$  and $z_{\rm max}=200$, respectively, and summing the powers of the first eight harmonics in both cases. The parameter $z_{\max}$ is defined as $T_{\rm obs}^2a_l/(cP)$, where $T_{\rm obs}$ is the observation duration, $a_l$ represents the line-of-sight acceleration caused by orbital motion, and $c$ denotes the speed of light. To ensure sensitivity to orbital periods as short as approximately 1.6\,hr, a ``segmented search'' approach was also employed. This involved dividing the observations, whenever possible, into sections of 120, 60, 30, and 10\,min, with each segment individually subjected to the above search process, giving a total of 3057 segments, requiring a total of approximately 4000\,hr of computing time and resources that included the Galahad and Hydrus computers at the University of Manchester. 

Additionally, we employed the technique known as `jerk search' \citep{Andersen+2018} implemented in \PRESTO, which enhanced our sensitivity to tight binary systems. This method focuses on identifying linear changes in acceleration ($\dot a$) within the Fourier spectrum by a number of Fourier bins determined by $w= \dot a T_{\rm obs}^3 /(cP)$. In our study, we extended the search up to a maximum value of $w_{\rm max}=600$ only for the full duration of the central beam of each observation. This limitation was imposed due to the significantly higher computational requirements of the technique. Notably, the concentration of pulsars within most globular clusters is typically higher towards its central region.

We kept only the candidates with a Fourier significance higher than 4$\sigma$ (for details see \citealt{Ransom+2002}) for visual inspection. Once an interesting candidate was found, we folded the neighbouring beams and the beams of the other observations at the interesting values of period, DM and acceleration. In order to confirm a candidate, we checked if it showed up in beams pointing in the same direction from the other observations.


\section{Discoveries}
\label{s:Discoveries}

The new pulsars discovered in this work are described in this section.

\subsection{M62H}
\label{s:Discovery_M62H}

\begin{figure*}
\centering
	\includegraphics[width=\textwidth]{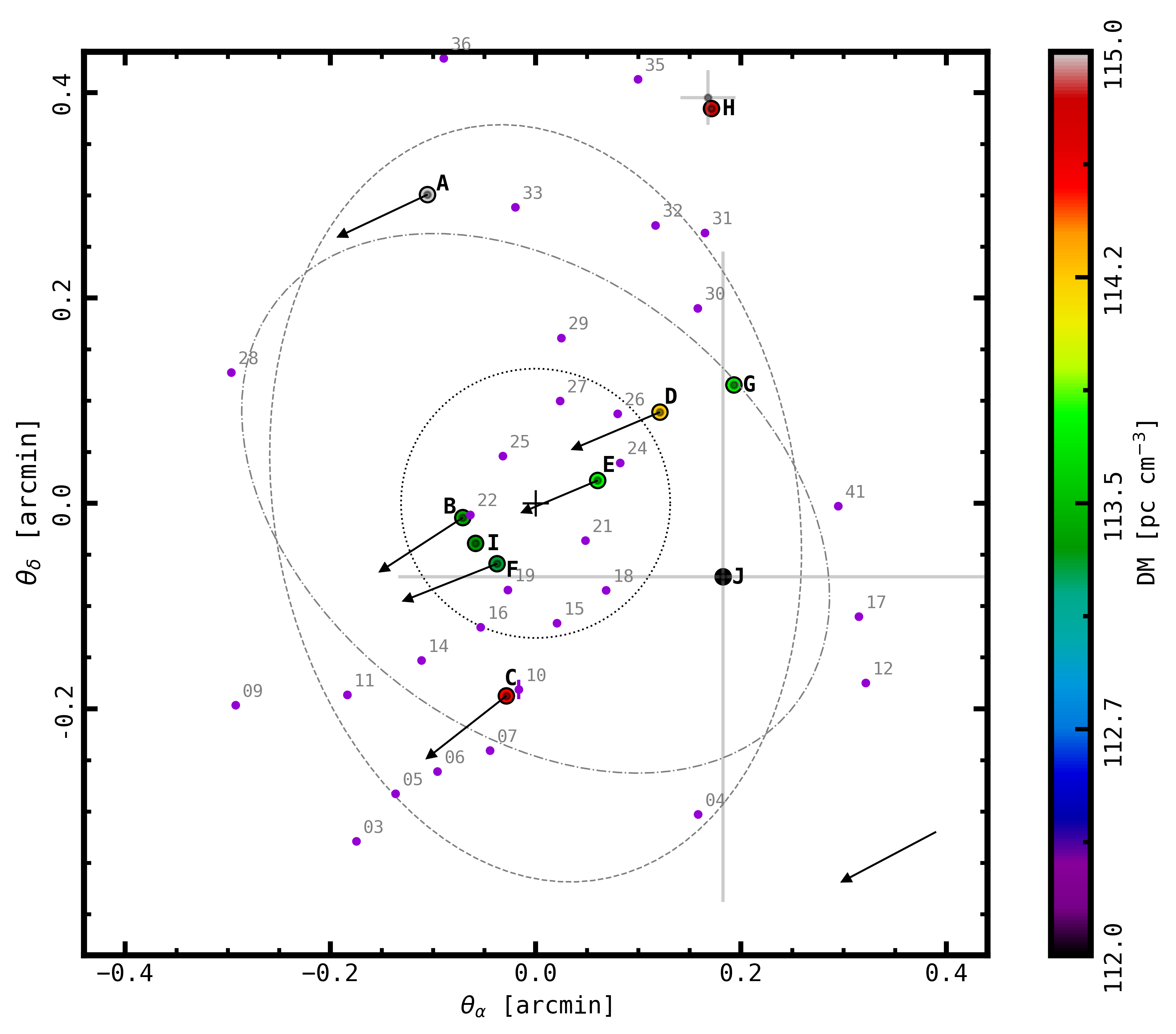}
    \caption{Positions of all the pulsars in M62, plotted as east-west ($\theta_{\alpha}$) and south-north ($\theta_{\delta}$) offsets from the centre of the GC. The grey crosses from pulsars M62H and M62J represent the 2$\sigma$ uncertainties in their positions obtained using \SeeKAT. The different colours of the symbols indicate their DM values. The angular core radius ($r_{\rm c} = 0.13$\,arcmin) is indicated by the dotted circle. We also show the sizes of both the L-band (dash-dotted ellipse) and UHF (dashed ellipse) beams for 56 antennas at their 50 per cent power level. The X-ray sources from \citet{Oh+2020} are shown in purple. The arrows show the direction and distance travelled by the pulsars in 1000\,yr considering the value of their proper motions. The proper motion of M62 from \citet{Baumgardt+Vasiliev2021} is shown in the bottom right corner.}
    \label{fig:M62_psrpositions}
\end{figure*}

M62H (also known as NGC~6266H or PSR~J1701--3006H) is a binary pulsar with a spin period of 3.70\,ms that was first found in the data from the beam 028 of the observation on UTC 2020-10-14-15:03 at a DM of 114.70\,pc\,cm$^{-3}$ with an acceleration of 0.288(5)\,m\,s$^{-2}$. It was found in the full, 2\,hr, 1\,hr, and 1/2\,hr segments of the \PRESTO~search. It was also detected in 7 beams of the same observation and later confirmed in an hour-long GBT observation from June 2009. This observation was taken using the GUPPI backend and the S-band (i.e. 1600-2400\,MHz) receiver. Unfortunately, this was the only detection we obtained with the GBT, so no GBT data is included in further analysis. After its confirmation, we pointed 12 beams around the highest S/N beam detection with a 90\% overlap in the observation on UTC 2021-12-30-12:27 in order to obtain a precise localisation. We then used the \SeeKAT~multibeam localiser\footnote{\url{https://github.com/BezuidenhoutMC/SeeKAT}} \citep{Bezuidenhout+2023} and the maximum likelihood localisation was found at coordinates $\alpha = 17\h01\m13\fs8(1)$ and $\delta =-30\degr06\arcmin24\arcsec(2)$. The grey cross in Figure \ref{fig:M62_psrpositions} is centred on the maximum likelihood position of the pulsar and the length of the cross corresponds to the reported errors, obtained from the 2$\sigma$ extent of the likelihood region.

We used the Period-Acceleration Diagram method (see \citealt{2001MNRAS_F}) to obtain starting estimates of the orbital parameters. First, we measured the barycentric observed spin period \Pobs and observed spin period derivative $\dot{P}_{\rm obs}$ from the five orbital campaign TRAPUM observations (see Table \ref{t:obs_summary}). We split the detections into 30\,min segments, giving us a total of twenty different measurements of \Pobs and $\dot{P}_{\rm obs}$. The $\dot{P}_{\rm obs}$ was then converted into the line-of-sight acceleration, $a_l = c(\dot{P}/P)_{\rm obs}$. This gave us first guess parameters of the orbit with an orbital period of $P_{\rm b} \simeq 2.7$\,hr, and a projected semi-major axis of only $x_{\rm p} \simeq 0.003$\,lt-s\footnote{Values obtained using the code \url{https://github.com/lailavc/circorbit}}. We then used \texttt{fit\_circular\_orbit.py} from \PRESTO, returning improved orbital parameters with an orbital period of $P_{\rm b} \simeq 3.2$\,hr and a projected semi-major axis of $x_{\rm p} \simeq 0.005$\,lt-s after fitting the observed spin period as a function of time, $P_{\rm obs}(t)$. Those values were further refined initially by phase-connecting the pulse times of arrival (ToA) from the observations of the orbital campaign (see Section \ref{s:Obs and Data}), and then obtaining a phase-connected timing solution for all the MeerKAT data using the \DRACULA\footnote{\url{https://github.com/pfreire163/Dracula}} algorithm \citep{Freire+Ridolfi2018}. See Section \ref{s:M62H}.

\subsection{M62I}

M62I (NGC~6266I or PSR~J1701--3006I) is a binary pulsar with a spin period of 3.30\,ms that was first found in the central beam of the observation on UTC 2021-09-21-15:55 at a DM of 113.35\,pc\,cm$^{-3}$ and an acceleration of 1.18(8)\,m\,s$^{-2}$. It was found in the 1\,hr, 30\,min, and 10\,min segments of the \PRESTO~search on UTC 2021-09-21-15:55, and confirmed in the observation on UTC 2021-12-30-12:27. We used the Period-Acceleration Diagram method to obtain starting estimates of the orbital parameters. This time the five observations from the orbital campaign were also split into 30\,min segments. Following the same procedure as for M62H, we obtain the first guess parameters of the orbital period $P_{\rm b} \simeq 12.1$\,hr and a projected semi-major axis of $x_{\rm p} \simeq 0.73$\,lt-s. These parameters were confirmed using \texttt{fit\_circular\_orbit.py}, obtaining an orbital period of $P_{\rm b} \simeq 12.2$\,hr and a projected semi-major axis of $x_{\rm p} \simeq 0.73$\,lt-s. Those values were then further refined by using timing. The results are presented in Section \ref{s:timing_M62I}. 

\subsection{M62J}

M62J (NGC~6266J or PSR~J1701--3006J) is a binary pulsar with a spin period of 2.76\,ms. It was initially found as a good candidate in one of the beams of the first UHF observation (UTC 2022-04-20-23:13) and later confirmed in the follow-up UHF observation (UTC 2022-12-01-14:20) at a DM of 111.98\,pc\,cm$^{-3}$ with an acceleration of 0.19(6)\,m\,s$^{-2}$. Since it was found in one of the edge beams we searched, we then carried out a PRESTO search of the neighbouring beams using the period, period derivative and DM from the highest S/N detection of the pulsar, and covering a range of 1$\times 10^{-4}$\,ms in period, and $\pm$0.25\,pc\,cm$^{-3}$ in DM, in steps of 0.05\,pc\,cm$^{-3}$ around the DM of 112.00\,pc\,cm$^{-3}$ for 10\,min, 30\,min, 1\,hr, and the full observation segments. As a result, the pulsar was detected in an additional three beams of the second UHF observation. These detections were later used for its localisation using \SeeKAT. The maximum likelihood localisation was found at coordinates $\alpha = 17\h01\m13\fs84(2)$ and $\delta = -30\degr06\arcmin52\arcsec(19)$. The grey cross shown in Figure \ref{fig:M62_psrpositions} is centred on this position and the length of the cross corresponds to the reported 2$\sigma$ errors. 

One beam was pointed towards the maximum likelihood position of the pulsar for each of the five TRAPUM orbital campaign observations. Six other beams were pointed around that position. We conducted blind searches of the pulsar following a similar procedure as for the neighbouring beams, using the same period and DM range for 10\,min, 30\,min, 1\,hr, and full observation segments. Since the bottom part of the band overlaps with the frequency range of the UHF observations, we repeated the same search procedure using only the first half of the band (962-1284\,MHz). Additionally, we searched for the pulsar by folding the data also using the pulsar period, period derivative and DM of our highest S/N detection. However, despite all of the different search methods employed, the pulsar was not detected in any of the L-band observations. 


\section{Timing}
\label{s:Timing}



In this section, we present the results of the long-term timing analysis conducted on all previously known pulsars within the globular cluster, along with the MeerKAT timing solutions for three of the recently discovered pulsars: M62G, M62H, and M62I.

Pulsars M62A-F all have timing solutions published by other authors \citep{Possenti+2003, Lynch+2012}. For each of these pulsars, we folded the Parkes ``Murriyang'' data at the topocentric period with the software package {\tt DSPSR}\footnote{\url{http://dspsr.sourceforge.net}}  \citep{vanStraten+2011}, as predicted by the ephemeris published in paper\footnote{Although we had access to these GBT data, they stopped in 2012 and overlapped with the Parkes data, and so we did not include them in the analysis presented here.} \cite{Lynch+2012}. We folded the data with a typical sub-integration length of 1\,min and maintained the same number of frequency channels as the raw data to accurately correct for the frequency-dependent signal dispersion due to the ionised interstellar medium.

We extracted the pulse ToAs with the routines of the software suite {\tt PSRCHIVE}\footnote{\url{http://psrchive.sourceforge.net/index.shtml}} \citep{Hotan+2004}. We coherently added in phase the profiles in each archive with respect to sub-integrations and channels, and convolved them with a high signal--to--noise (S/N) template, that we built by summing in phase the observed profiles with the highest S/N.

We performed the ToAs extraction on a per-observation basis, in order to obtain, from each archive, the largest reasonable number of ToAs. We thus tried different decimation schemes either in time only, frequency only, or both, and maintained the one that allowed us to obtain the highest number of ToAs after applying the following procedure. In correspondence with each decimation scheme, we visually inspected the obtained profiles, and accepted the ToAs whose corresponding profile were evaluated as detected and not too degraded with respect to the profile obtained by coherently summing all subintegrations and all channels together. This procedure allowed us to obtain more than one ToA from the majority of observations.

We used the obtained ToAs to derive a preliminary ephemeris for pulsars M62A-F, and we used them to repeat the whole procedure mentioned above. At the end of this second iteration we obtained the ToAs we used in this work. In the case of pulsar M62B, we also added ToAs obtained from the Effelsberg observations. 

We then used the ephemeris derived from the Parkes ``Murriyang'' (and Effelsberg for the case of M62B) data to fold the MeerKAT search-mode data using \texttt{DSPSR} for the previously known pulsars. Pulsar M62G was reported to have a partial-phase connected solution \citep{Ridolfi+2021}. We then used the ephemeris published in \citealt{Ridolfi+2021} to fold the MeerKAT data as our initial ephemeris. 

Regarding the new discoveries (M62H and M62I), we also folded the data using \texttt{DSPSR} and an initial ephemeris that included the barycentric spin period and DM of the best detection obtained with \texttt{prepfold} routine from \PRESTO, and the orbital parameters from \texttt{fit\_circular\_orbit.py}. We then utilised the \texttt{pat} tool from the \texttt{PSRCHIVE} package and a noise-free template to extract the topocentric ToAs from each archive in which the pulsar was detected. 

The template was constructed by fitting von Mises functions to either the highest S/N profile or, where possible, one formed by combining observations with detections to get a higher S/N. Different standard profiles were employed for the simultaneous MeerTime and TRAPUM data (see Section \ref{s:Obs and Data}). Subsequently, a precise measurement of the time offset between the two datasets was obtained by using the simultaneous observations of the two different MeerKAT backends (PTUSE and APSUSE).

\begin{figure*}
	\includegraphics[width=\textwidth, height=15cm]{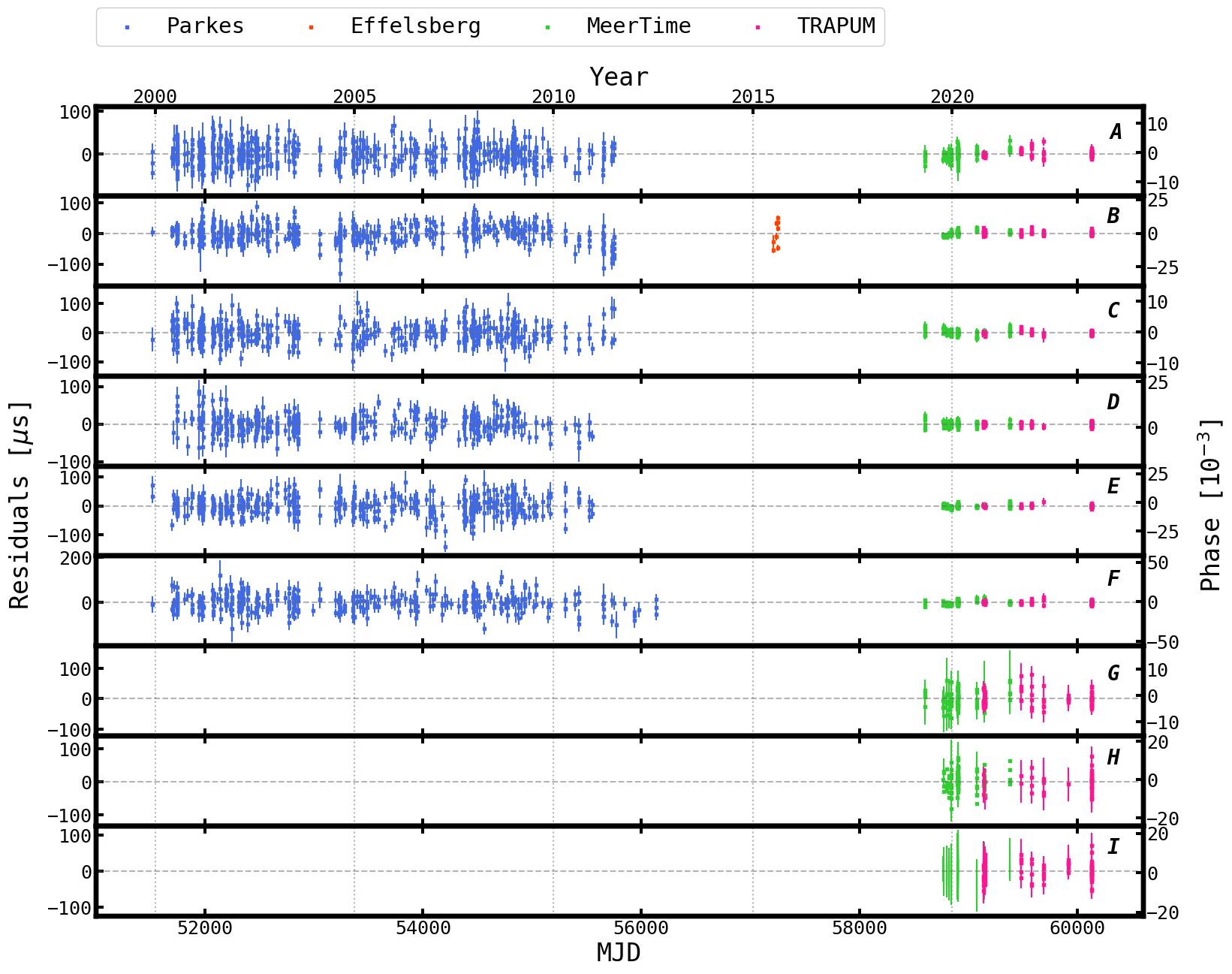}
    \caption{Timing residuals as a function of time obtained using Parkes, Effelsberg and MeerKAT data for all the previously known pulsars using a data span of over 23\,yr (M62A--F), and for the newly discovered pulsars with only MeerKAT solutions (M62--M62I) using a data span of over 3\,yr. The blue points indicate times of arrival from Parkes, the orange are from Effelsberg, while the green and pink points show the times of arrival from MeerTime and TRAPUM, respectively.}
        \label{fig:timing_mjd_plots}
\end{figure*}

The ToAs were then converted to the Solar System Barycentre reference and fitted for different timing model parameters using \texttt{TEMPO2}\footnote{\url{https://bitbucket.org/psrsoft/tempo2/src/master/}} \citep{Hobbs+2006} using the Jet Propulsion Laboratory  (JPL) DE421 Solar System ephemeris \citep{Folkner+2009}. These parameters encompassed celestial coordinates, spin parameters, and orbital parameters. The timing solutions of the non-eclipsing pulsars are presented in alphabetical order in Tables \ref{tab:timing_fitresults_1} and \ref{tab:timing_fitresults_2}. The solutions of the two eclipsing pulsars and pulsar M62H are presented in Table \ref{tab:timing_fitresults_3}. The outcomes of the timing analysis for each pulsar are presented in the subsequent subsections.

\begin{figure*}
	\includegraphics[width=\textwidth, height=15cm]{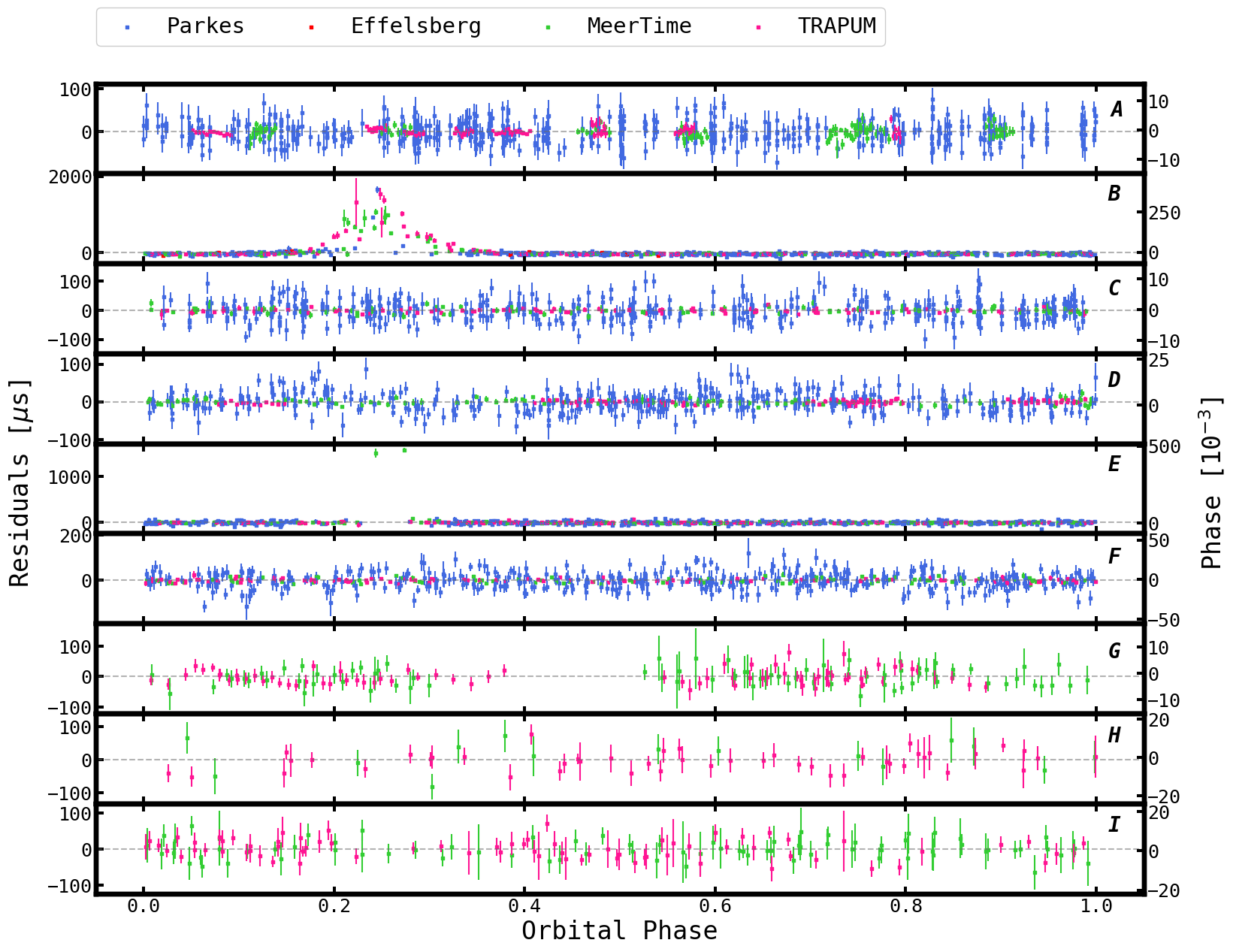}
    \caption{Timing residuals as a function of orbital phase obtained using Parkes, Effelsberg and MeerKAT data for all the previously known pulsars (M62A--F), and for the newly discovered pulsars with only MeerKAT solutions (M62G--I). The ToAs for pulsars M62B and M62E show delays around superior conjunction, these are caused by eclipses. The blue points indicate times of arrival from Parkes, the orange are from Effelsberg, while the green and pink points show the times of arrival from MeerTime and TRAPUM, respectively.}
        \label{fig:timing_orbphase_plots}
\end{figure*}

\subsection{Pulsars with previously published timing solutions}
\label{s:timing_knownpsrs}
For the pulsars with a previously published timing solution, we find that nearly all our measured parameters agree with those published elsewhere to within errors (with some exceptions, see below), and the errors on our measurements are up to 10 times smaller. Moreover, we were able to measure, for the first time, the second spin-period derivatives, the proper motions, and the changes in the orbital period for all these pulsars, and the third spin-period derivatives in some particular cases. A detailed interpretation of these spin-period derivative measurements is presented in section \ref{s:Discussion}. Figures \ref{fig:timing_mjd_plots} and \ref{fig:timing_orbphase_plots} show the timing residuals as a function of MJD and orbital phase, respectively, for these pulsars.

\begin{table*}
\setlength\tabcolsep{10pt}
  \begin{threeparttable}
    \caption{Timing parameters for the pulsars M62A, M62C and M62D, as obtained from fitting the observed ToAs with \TEMPOTWO. The companion mass is calculated assuming a pulsar mass of 1.4 \msun. The time units are TDB, the adopted terrestrial time standard is TT(BIPM2019) and the Solar System ephemeris used is JPL DE421 \citep{Folkner+2009}. Numbers in parentheses represent 1-$\sigma$ uncertainties in the last digit. The DM values were obtained using \texttt{PDMP}.}
    \centering
     \label{tab:timing_fitresults_1}
    \renewcommand{\arraystretch}{1.0}
     \begin{tabular}{l c c c} 
     \hline
      \hline
Pulsar  &  M62A  &  M62C  &  M62D  \\
\hline
Right Ascension, $\alpha$ (hh:mm:ss.s; J2000) \dotfill   &  17:01:12.50899(5)  & 17:01:12.86420(6) & 17:01:13.55601(4) \\
Declination, $\delta$ (dd:mm:ss.s; J2000) \dotfill   & $-$30:06:30.168(4)  & $-$30:06:59.457(5) & $-$30:06:42.877(4) \\
Proper Motion in $\alpha$, $\mu_\alpha \cos \delta$ (mas\,yr$^{-1}$) \dotfill &  $-$5.31(7) & $-$4.72(8) & $-$5.20(7) \\
Proper Motion in $\delta$, $\mu_\delta$ (mas\,yr$^{-1}$) \dotfill &  $-$2.6(5) & $-$3.7(6) & $-$2.2(5) \\
Spin Frequency, $f$ (s$^{-1}$)  \dotfill 	& 190.782671214183(7) & 131.356872901228(9) & 292.588403183803(14) \\
1st Spin Frequency derivative, $\dot{f}$ (Hz\,s$^{-1}$) \dotfill & 4.793645(17) $\times 10^{-15}$ & 1.09266(6)$\times 10^{-15}$ & $-$1.161178(3)$\times 10^{-14}$ \\
2nd Spin Frequency derivative, $\ddot{f}$ (Hz\,s$^{-2}$) \dotfill & $-$7.22(20)$\times 10^{-27}$ & $-$8.1(3)$\times 10^{-27}$ & $-$6.17(4)$\times 10^{-26}$ \\
3rd Spin Frequency derivative, $f^{(3)}$ (Hz\,s$^{-3}$) \dotfill & -- & 4.5(4)$\times 10^{-35}$ & -- \\
Reference Epoch (MJD)  \dotfill & 55754.676 & 55754.669 &  55550.225 \\
Start of Timing Data (MJD) \dotfill & 51521.005 & 51521.042 & 51714.717 \\
End of Timing Data (MJD) \dotfill 	&  60130.786 &  60131.035 & 60131.035 \\
Dispersion Measure, DM (pc\,cm$^{-3}$) \dotfill	& 114.98(4) & 114.55(5) & 114.23(2) \\
Number of ToAs \dotfill     & 813 & 674 & 618 \\
Weighted r.m.s. residual ($\mu$s) \dotfill  &  12.2 & 14.6 & 10.2 \\
$S_{816}$ (mJy) \dotfill & 0.43 & 0.19 & 0.85 \\
$L_{816}$ (mJy kpc$^2$) \dotfill  & 29.97 & 13.42 & 58.89 \\
$S_{1284}$ (mJy) \dotfill & 0.13 & 0.03 & 0.14 \\
$L_{1284}$ (mJy kpc$^2$) \dotfill  & 8.75 & 2.31 & 9.44 \\
Spectral Index, $\beta$ \dotfill  & -2.71 & -3.89 & -4.04 \\
\hline
\multicolumn{4}{c}{Binary Parameters}  \\
\hline
Binary Model \dotfill   & ELL1 & ELL1 & ELL1 \\
Orbital Period, $P_{\rm b}$ (days) \dotfill &  3.8059483909(12) & 0.21500007182(8) & 1.11790340055(11) \\
1st Orbital Period derivative \dotfill  & $-$1.32(3)$\times 10^{-11}$ & $-$2.76(18)$\times 10^{-13}$ & 3.37(8)$\times 10^{-12}$ \\
Projected Semi-major Axis, $x_{\rm {p}}$ (lt-s)  \dotfill &  3.4837265(9) & 0.1928888(9) & 0.9880515(8) \\
Epoch of Ascending Node, $T_\textrm{asc}$ (MJD) \dotfill &  52048.5627983(4) & 52049.8556544(6) & 55037.8870645(4) \\

1st Laplace-Lagrange parameter, $\epsilon = e \sin \omega$ \dotfill & 4.6(5)$\times 10^{-6}$ & 3.0(9)$\times 10^{-5}$ & 4.366(16)$\times 10^{-4}$ \\
2st Laplace-Lagrange parameter, $\epsilon = e \cos \omega$ \dotfill & $-$4.8(5)$\times 10^{-6}$ & $-$1.4(9)$\times 10^{-5}$ & $-$1.344(15)$\times 10^{-4}$ \\
\hline
\multicolumn{4}{c}{Derived Parameters}  \\
\hline
Spin Period, $P$ (ms)   \dotfill & 5.2415661948529(2) &  7.6128487068349(5) & 3.4177704554196(2) \\ 
1st Spin Period derivative, $\dot{P}$ (s\,s$^{-1}$)  \dotfill & $-$1.317006(5)$\times 10^{-19}$ & $-$6.3325(4)$\times 10^{-19}$ &  1.356389(3)$\times 10^{-19}$ \\
L.o.s. accel. from the cluster field, $a_{l,{\rm GC}}$ (m\,s$^{-2}$) \dotfill & $-$1.26(3)$\times10^{-8}$ & $-$0.50(3)$\times10^{-8}$ & 1.0(2)$\times10^{-8}$ \\
Intrinsic spin period derivative, $\dot{P}_{\rm int}$ (s\,s$^{-1}$) \dotfill & 7.87 $\times10^{-20}$ & 4.98$\times10^{-20}$ & 1.64$\times10^{-20}$ \\
Characteristic age$^{\dagger}$, $\tau_{\rm c}$ (yr) \dotfill & 1.05$\times 10^{9}$ & 2.42$\times 10^{9}$ & 3.29$\times 10^{9}$ \\
Surface magnetic field$^{\dagger}$, $B_{\rm s}$ (G) \dotfill & 6.50 $\times 10^{8}$  & 6.23 $\times 10^{8}$ & 2.40$\times 10^{8}$ \\
L.o.s. jerk, $\dot{a_l}$ (m\,s$^{-3}$), \dotfill & 1.13(3)$\times10^{-20}$ & 1.85(7)$\times10^{-20}$ & 6.32(4)$\times10^{-20}$ \\
Epoch of Periastron, $T_\textrm0$ (MJD) \dotfill &  52050.00(4) &   52049.924(8) &  55038.2196(6) \\
Eccentricity, $e$ \dotfill & 6.4(5) $\times 10^{-6}$  &  3.4(9)$\times 10^{-5}$ & 456(2) $\times 10^{-6}$ \\
Longitude of periastron, $\omega$ ($\deg$) \dotfill & 122(4) &  114(14) & 107.1(2) \\
Mass Function, $f(M_{\rm p})$ (\msun)   \dotfill &  0.00313393(2) &  1.66696(2)$\times 10^{-4}$ & 8.28729(2)$\times 10^{-4}$ \\
Minimum companion mass, $M_{\rm c, min}$ (\msun) \dotfill & 0.1956 & 0.06951 & 0.1215 \\
Median companion mass, $M_{\rm c, med}$ (\msun)  \dotfill & 0.2291 & 0.0807 & 0.1416 \\
    \hline
      \hline
     \end{tabular}
\begin{tablenotes}
      \small
      \item[${\dagger}$] Calculated using $\dot{P}_{\rm int}$.
\end{tablenotes}
  \end{threeparttable}
\end{table*}

\begin{table*}
\setlength\tabcolsep{10pt}
  \begin{threeparttable}
    \caption{Timing parameters for the pulsars M62F, M62G and M62I, as obtained from fitting the observed ToAs with \TEMPOTWO. The companion mass is calculated assuming a pulsar mass of 1.4 \msun. The time units are TDB, the adopted terrestrial time standard is TT(BIPM2019) and the Solar System ephemeris used is JPL DE421 \citep{Folkner+2009}. Numbers in parentheses represent 1-$\sigma$ uncertainties in the last digit. The DM values were obtained using \texttt{PDMP}.}
    \centering
     \label{tab:timing_fitresults_2}
    \renewcommand{\arraystretch}{1.0}
     \begin{tabular}{l c c c} 
     \hline
      \hline
Pulsar  & M62F & M62G & M62I\\
\hline
Right Ascension, $\alpha$ (hh:mm:ss.s; J2000) \dotfill  & 17:01:12.82232(6) &  17:01:13.8892(3) & 17:01:12.72568(17) \\
Declination, $\delta$ (dd:mm:ss.s; J2000) \dotfill  &  $-$30:06:51.734(5) &  $-$30:06:41.265(18) & $-$30:06:50.55(2) \\
Proper Motion in $\alpha$, $\mu_\alpha \cos \delta$ (mas\,yr$^{-1}$) \dotfill & -5.56(9) & -- & -- \\
Proper Motion in $\delta$, $\mu_\delta$ (mas\,yr$^{-1}$) \dotfill &  -2.2(6) & -- & -- \\
Spin Frequency, $f$ (s$^{-1}$)  \dotfill 	& 435.781667403045(5) & 217.00905523505(4) & 303.43334924566(3) \\
1st Spin Frequency derivative, $\dot{f}$ (Hz\,s$^{-1}$) \dotfill & $-$4.192540(3)$\times 10^{-14}$ &  5.3952(10)$\times 10^{-15}$ & 3.10515(12)$\times 10^{-14}$ \\
2nd Spin Frequency derivative, $\ddot{f}$ (Hz\,s$^{-2}$) \dotfill &  $-$2.733(3)$\times 10^{-25}$ & -- & --\\
Reference Epoch (MJD)  \dotfill & 55848.40 & 59377.84 & 59478.67\\
Start of Timing Data (MJD) \dotfill & 51521.006 & 58602.827 & 58769.446  \\
End of Timing Data (MJD) \dotfill 	&  60131.035 & 60131.032 & 60131.035\\
Dispersion Measure, DM (pc\,cm$^{-3}$) \dotfill	& 113.29(2) & 113.67(3) & 113.35(2) \\
Number of ToAs \dotfill  &  605 & 162& 188 \\
Weighted r.m.s. residual ($\mu$s) \dotfill  &  13.6 & 22.5 & 35.7 \\
$S_{816}$ (mJy) \dotfill & 0.22 &  0.10 & 0.04 \\
$L_{816}$ (mJy kpc$^2$) \dotfill  & 15.39 & 7.08 & 2.60 \\
$S_{1284}$ (mJy) \dotfill & 0.11 &  0.03 & 0.02 \\
$L_{1284}$ (mJy kpc$^2$) \dotfill  & 7.58 & 2.13 & 1.31 \\
Spectral Index, $\beta$ \dotfill  & -1.56 & -2.65 & -1.51 \\
\hline
\multicolumn{4}{c}{Binary Parameters}  \\
\hline
Binary Model \dotfill   & ELL1 & ELL1 & ELL1 \\
Orbital Period, $P_{\rm b}$ (days) \dotfill &  0.2054870463(1) & 0.774433617(1) & 0.509252794(1) \\
1st Orbital Period derivative \dotfill  & 1.23(5)$\times 10^{-12}$ & -- & $-$7(2)$\times 10^{-12}$\\
Projected Semi-major Axis, $x_{\rm {p}}$ (lt-s)  \dotfill &  0.0573449(9) & 0.620319(3) & 0.732802(4)\\
Epoch of Ascending Node, $T_\textrm{asc}$ (MJD) \dotfill & 58590.1382864(8) & 58894.9663224(14) & 60131.0203640(6) \\
1st Laplace-Lagrange parameter, $\epsilon = e \sin \omega$ \dotfill &  6(2)$\times 10^{-5}$ & 5.8(2)$\times 10^{-4}$ & 2.00(9)$\times 10^{-4}$ \\
2st Laplace-Lagrange parameter, $\epsilon = e \cos \omega$ \dotfill &  -3(30) & 6.71(1)$\times 10^{-4}$ & $-$5(1)$\times 10^{-5}$ \\

\hline
\multicolumn{4}{c}{Derived Parameters}  \\
\hline
Spin Period, $P$ (ms)   \dotfill &  2.2947270957020(2) & 4.6081026384677(4) & 3.2956166369462(7) \\ 
1st Spin Period derivative, $\dot{P}$ (s\,s$^{-1}$)  \dotfill &  2.207696(2)$\times 10^{-19}$ & $-$1.1457(2)$\times 10^{-19}$ & $-$3.3725(1) \\
L.o.s. accel. from the cluster field, $a_{l,{\rm GC}}$ (m\,s$^{-2}$) \dotfill & 2.02(8) & -- & $-$5(1) \\
Intrinsic spin period derivative, $\dot{P}_{\rm int}$ (s\,s$^{-1}$) \dotfill & 6.18$\times10^{-20}$ & -- & 1.87$\times10^{-19}$ \\
Characteristic age$^{\dagger}$, $\tau_{\rm c}$ (yr) \dotfill & 5.87$\times 10^{8}$ & --& >2.78$\times 10^{8}$ \\
Surface magnetic field$^{\dagger}$, $B_{\rm s}$ (G) \dotfill & 3.81 $\times 10^{8}$ & -- & <7.95$\times 10^{8}$ \\
L.o.s. jerk, $\dot{a_l}$ (m\,s$^{-3}$), \dotfill & 1.880(5)$\times10^{-19}$ & -- & -- \\
Epoch of Periastron, $T_\textrm{0}$ (MJD) \dotfill & 58590.19(1) & 58895.053(2) & 60131.170(3) \\
Eccentricity, $e$ \dotfill & 7(3)$\times 10^{-5}$ & 8.8(2)$\times 10^{-5}$ & 2.08(9)$\times 10^{-4}$\\
Longitude of periastron, $\omega$ ($\deg$) \dotfill & 92(25) & 40.7(8) & 105(2)\\
Mass Function, $f(M_{\rm p})$ (\msun)  \dotfill &  4.7951(2)$\times 10^{-6}$ & 4.2723(6)$\times 10^{-4}$ & 1.62921(2)$\times 10^{-3}$ \\
Minimum companion mass, $M_{\rm c, min}$ (\msun) \dotfill & 0.02081 & 0.09633 & 0.1545\\
Median companion mass, $M_{\rm c, med}$ (\msun)  \dotfill & 0.0241 & 0.1120 & 0.1804\\
    \hline
      \hline
     \end{tabular}
\begin{tablenotes}
      \small
      \item[${\dagger}$] Calculated using $\dot{P}_{\rm int}$ where it was available.
\end{tablenotes}
  \end{threeparttable}
\end{table*}

\begin{table*}
\setlength\tabcolsep{10pt}
  \begin{threeparttable}
    \caption{Timing parameters for the two eclipsing systems in M62 as obtained from fitting the observed ToAs with \TEMPOTWO, and for M62H as obtained using \DRACULA. The companion mass is calculated assuming a pulsar mass of 1.4 \msun. The time units are TDB, the adopted terrestrial time standard is TT(BIPM2019) and the Solar System ephemeris used is JPL DE421 \citep{Folkner+2009}. Numbers in parentheses represent 1-$\sigma$ uncertainties in the last digit. The DM values were obtained using \texttt{PDMP}.}
    \centering
     \label{tab:timing_fitresults_3}
    \renewcommand{\arraystretch}{1.0}
     \begin{tabular}{l c c c} 
     \hline
      \hline
Pulsar  &   M62B  & M62E & M62H \\
\hline
Right Ascension, $\alpha$ (hh:mm:ss.s; J2000) \dotfill   &  17:01:12.66786(5)   & 17:01:13.27554(4) & 17:01:13.7881(5) \\
Declination, $\delta$ (dd:mm:ss.s; J2000) \dotfill   & $-$30:06:49.039(4)   & $-$30:06:46.869(4) & $-$30:06:25.14(5) \\
Proper Motion in $\alpha$, $\mu_\alpha \cos \delta$ (mas\,yr$^{-1}$) \dotfill &  $-$4.92(7)  & $-$4.51(6) & -- \\
Proper Motion in $\delta$, $\mu_\delta$ (mas\,yr$^{-1}$) \dotfill &  $-$3.2(5) & $-$1.9(4) & -- \\
Spin Frequency, $f$ (s$^{-1}$)  \dotfill & 278.252964025804(15) & 309.239709508853(3) & 269.92298522974(6)  \\
1st Spin Frequency derivative, $\dot{f}$ (Hz\,s$^{-1}$) \dotfill & 2.706912(10)$\times 10^{-14}$  & $-$2.9400470(13)$\times 10^{-14}$ & $-$2.760(3)$\times 10^{-15}$\\
2nd Spin Frequency derivative, $\ddot{f}$ (Hz\,s$^{-2}$) \dotfill &  $-$5.9(3)$\times 10^{-27}$  & 4.568(14)$\times 10^{-26}$ & -- \\
3rd Spin Frequency derivative, $f^{(3)}$ (Hz\,s$^{-3}$) \dotfill & 8.5(6)$\times 10^{-35}$ & --  & -- \\
Reference Epoch (MJD)  \dotfill & 55754.685  & 55550.236 &  59478.677 \\
Start of Timing Data (MJD) \dotfill & 51521.042  & 51521.006 & 58769.490 \\
End of Timing Data (MJD) \dotfill 	&  60131.036  & 60131.036  & 60131.032 \\
Dispersion Measure, DM (pc\,cm$^{-3}$) \dotfill	& 113.46(3)  & 113.79(2) & 114.70(3)  \\
Number of ToAs \dotfill     & 713  & 759 & 62 \\
Weighted r.m.s. residual ($\mu$s) \dotfill  &  11.9  & 7.5 & 30.3\\
$S_{816}$ (mJy) \dotfill & 0.78 &  0.88 & 0.05 \\
$L_{816}$ (mJy kpc$^2$) \dotfill  & 53.48 & 60.62 & 3.39 \\
$S_{1284}$ (mJy) \dotfill & 0.47 &  0.23 & 0.05 \\
$L_{1284}$ (mJy kpc$^2$) \dotfill  & 32.04 & 16.17 & 3.62 \\
Spectral Index, $\beta$ \dotfill  & -1.13 & -2.92 & 0.15 \\
\hline
\multicolumn{4}{c}{Binary Parameters}  \\
\hline
Binary Model \dotfill   & BTX  & BTX & BT \\
Orbital Period, $P_{\rm b}$ (days) \dotfill & 0.144545416(1) & 0.158477359(3) & 0.132935028(8) \\
Orbital Frequency, $f_{\rm b}$ (s$^{-1}$) \dotfill & 8.007223182(6)$\times 10^{-5}$ & 7.303298167(14)$\times 10^{-5}$ & -- \\
1st Orbital Frequency derivative, $f_{\rm b}^{(1)}$ (s$^{-2}$)  \dotfill  & 4.80(9)$\times 10^{-20}$  & 7(1)$\times 10^{-20}$ & -- \\
2nd Orbital Frequency derivative, $f_{\rm b}^{(2)}$ (s$^{-3}$) \dotfill  & 1.4(3)$\times 10^{-28}$  & 1.2(2)$\times 10^{-26}$ & -- \\
3rd Orbital Frequency derivative, $f_{\rm b}^{(3)}$ (s$^{-4}$) \dotfill  & $-$1.4(6)$\times 10^{-36}$  &  3(3)$\times 10^{-35}$ & -- \\
4th Orbital Frequency derivative, $f_{\rm b}^{(4)}$ (s$^{-5}$) \dotfill  & $-$5(2)$\times 10^{-44}$  & $-$1.9(1)$\times 10^{-41}$ & -- \\
5th Orbital Frequency derivative, $f_{\rm b}^{(5)}$ (s$^{-6}$)  \dotfill  & 2.4(2.3)$\times 10^{-52}$  & $-$8.4(10)$\times 10^{-49}$ & -- \\
6th Orbital Frequency derivative, $f_{\rm b}^{(6)}$ (s$^{-7}$) \dotfill  & 2.5(9)$\times 10^{-59}$  & $-$2.0(3)$\times 10^{-56}$ & -- \\
7th Orbital Frequency derivative, $f_{\rm b}^{(7)}$ (s$^{-8}$) \dotfill  & $-$1.8(7)$\times 10^{-67}$  & $-$3.1(6)$\times 10^{-64}$ & -- \\
8th Orbital Frequency derivative, $f_{\rm b}^{(8)}$ (s$^{-9}$) \dotfill  & $-$7(4)$\times 10^{-75}$  & $-$3.0(7)$\times 10^{-72}$ & -- \\
9th Orbital Frequency derivative, $f_{\rm b}^{(9)}$ (s$^{-10}$) \dotfill  & 10(4)$\times 10^{-83}$  & $-$1.8(6)$\times 10^{-80}$ & -- \\
10th Orbital Frequency derivative, $f_{\rm b}^{(10)}$ (s$^{-11}$) \dotfill & $--$  & $-$4(2)$\times 10^{-89}$ & -- \\
Projected Semi-major Axis, $x_{\rm {p}}$ (lt-s)  \dotfill &  0.2527621(9)  & 0.0573450(9) & 0.004908(6) \\
Epoch of Periastron, $T_\textrm{0}$ (MJD) \dotfill &  55000.0321376(4)  &  58590.1382864(8) & 60129.86532(5) \\
\hline
\multicolumn{4}{c}{Derived Parameters}  \\
\hline
Spin Period, $P$ (ms)   \dotfill & 3.59385212820244(3)  & 3.23373735406827(2) & 3.704760451395(2) 
 \\ 
1st Spin Period derivative, $\dot{P}$ (s\,s$^{-1}$)  \dotfill & -3.49618(2)$\times 10^{-19}$  & 3.074423(2)$\times 10^{-19}$ & 3.789(4)$\times 10^{-20}$ \\
L.o.s. jerk, $\dot{a_l}$ (m\,s$^{-3}$), \dotfill & 6(3)$\times10^{-21}$ & -4.43(1)$\times10^{-20}$  & -- \\
Mass Function, $f(M_{\rm p})$ (\msun)   \dotfill &  8.29849(8) $\times 10^{-4}$ & 1.47578(3)$\times 10^{-5}$ & 7.18(3)$\times 10^{-9}$ \\
Minimum companion mass, $M_{\rm c, min}$ (\msun) \dotfill & 0.1216  & 0.03041 & 0.00236 \\
Median companion mass, $M_{\rm c, med}$ (\msun)  \dotfill & 0.1417  & 0.0352 & 0.0027 \\
    \hline
      \hline
     \end{tabular}
  \end{threeparttable}
\end{table*}

\subsubsection{M62A}
\label{s: Timing_A}

M62A is a pulsar in a binary system with an orbital period of 3.8\,days with a minimum companion mass $M_{\rm c, min}=0.2$\,\msun. Our measured parameters agree with those previously published \citep{Possenti+2003,Lynch+2012}, except for the eccentricity. Using the ELL1 model \citep{Lange+2001}, we obtain an inferred eccentricity $e=6.4(5)\times 10^{-6}$, which is $\sim 3\sigma$ away from the previously estimated value. The MeerKAT sensitivity and the longer data set allow us to obtain, for the first time, measurements of the longitude of periastron passage $\omega = 122(4)$\,$\deg$ and the rate of change of orbital period $\dot{P}_{\rm b} =-1.32(3)\times 10^{-11}$. The long-term timing solution allows us to obtain measurements of the proper motion of the pulsar: $\mu_{\alpha} \cos \delta = -5.31(7)$\,mas\,yr$^{-1}$ and $\mu_{\delta} = -2.6(5)$\,mas\,yr$^{-1}$, as well as the second spin frequency derivative $\ddot{f} = -7.2(2) \times 10^{-27}$\,Hz\,s$^{-3}$.

\subsubsection{M62B}

Pulsar M62B is an eclipsing redback with an orbital period of $\sim 3.5$\,hr and a $M_{\rm c, min}=0.1$\,\msun. Previous published timing solutions for this pulsar used the ELL1 timing model, whereas in our case, we use a BTX model \citep{Shaifullah+2016} which allows a description of the orbital behaviour using nine orbital frequency derivatives (see Table \ref{tab:timing_fitresults_3}). For this model, both the eccentricity and the longitude of periastron passage are set to zero. Our measurement of the rate of change of the orbital period $\dot{P_{\rm b}} = -7.5(1)\times 10^{-12}$, is almost 4$\sigma$ away from the value reported in \citep{Lynch+2012}. This is likely because it is a non-stationary value, as evidenced by the large number of derivatives. The long-term timing solution also allows us to obtain measurements of the proper motion of the pulsar: $\mu_{\alpha}\cos \delta = -4.92(7)$\,mas\,yr$^{-1}$ and $\mu_{\delta} = -3.2(5)$\,mas\,yr$^{-1}$, the second spin frequency derivative $\ddot{f} = -7.2(2) \times 10^{-27}$\,Hz\,s$^{-2}$, and the third frequency derivative $f^{(3)} = 8.5(6) \times 10^{-35}$\,Hz\,s$^{-3}$.

\subsubsection{M62C}

M62C is a binary pulsar with an orbital period of $\sim 5.2$\,hr and with a light companion mass of $M_{\rm c, min}=0.07$\,\msun. Our measurement of the position $\alpha = 17\h01\m12\fs86420(6)$ and $\delta = -30\degr06\arcmin59\farcs457(5)$ is almost 5$\sigma$ away from the value previously published by \citet{Lynch+2012} in $\alpha$ and 2$\sigma$ in $\delta$. However, the angular offset between the two positions is of 0.04\,arcsec. We also obtain, for the first time for this pulsar, measurements of the longitude of periastron passage $\omega = 114(14)$\,$\deg$ and the rate of change of orbital period $\dot{P}_{\rm b} = -2.8(2)\times 10^{-13}$. The long-term timing solution also allows us to obtain measurements of the proper motion of the pulsar: $\mu_{\alpha} \cos \delta = -4.72(8)$\,mas\,yr$^{-1}$ and $\mu_{\delta} = -3.7(6)$\,mas\,yr$^{-1}$, the second spin frequency derivative $\ddot{f} = -8.1(3) \times 10^{-27}$\,Hz\,s$^{-2}$, and the third frequency derivative $f^{(3)} = 4.5(4) \times 10^{-35}$\,Hz\,s$^{-3}$.

\subsubsection{M62D}

M62D is a binary pulsar with an orbital period of 1.1\,days and a $M_{\rm c, min}= 0.1$\,\msun. Our measured values of the position $\alpha = 17\h01\m13\fs55601(6)$ and $\delta = -30\degr06\arcmin42\farcs877(4)$ are $\sim 5\sigma$ and $\sim 4 \sigma$ from the values previously published by \citet{Lynch+2012}, respectively. This corresponds to a 0.33\,arcsec angular offset between the two positions. Moreover, our estimate of the eccentricity ($e=4.56\times 10^{-6}$), longitude of periastron ($\omega = 107.2(2)$\,$\deg$), and projected semi-major axis ($x_{\rm p} = 0.9880515$(8)\,lt-s), are all $\sim 6\sigma$ away from the previous measured values. The long-term timing solution also allow us to obtain measurements of the proper motion of the pulsar: $\mu_{\alpha} \cos \delta = -5.20(7)$\,mas\,yr$^{-1}$ and $\mu_{\delta} = -2.2(5)$\,mas\,yr$^{-1}$, as well as the second spin frequency derivative $\ddot{f} = -6.17(4) \times 10^{-26}$\,Hz\,s$^{-2}$.

\subsubsection{M62E}

M62E is an eclipsing black widow with an orbital period of $\sim 3.8$\,hr and a light companion mass $M_{\rm c, min}= 0.03$\,\msun. For this pulsar, we also use a BTX binary model compared to an ELL1 from \citet{Lynch+2012}, making use of ten orbital frequency derivatives (see Table \ref{tab:timing_fitresults_3}). From this, we measure the rate of change of the orbital period $\dot{P_{\rm b}} = -1.3(2)\times 10^{-11}$. The long-term timing solution also allow us to obtain measurements of the proper motion of the pulsar: $\mu_{\alpha}\cos \delta = -4.51(6)$\,mas\,yr$^{-1}$ and $\mu_{\delta} = -1.9(4)$\,mas\,yr$^{-1}$, as well as the second spin frequency derivative $\ddot{f} = -4.57(2) \times 10^{-26}$\,Hz\,s$^{-2}$.

\subsubsection{M62F}

M62F is classified as a non-eclipsing black widow with an orbital period of $\sim 5$\,hr and light companion mass $M_{\rm c, min}= 0.02$\,\msun. There is no evidence of eclipses in any of the very sensitive MeerKAT observations, supporting the non-eclipsing nature of the source (see Section \ref{s:Discussion} for a discussion). We obtained the measurement of the rate of change of the orbital period $\dot{P}_{\rm b} = 1.23(5) \times 10^{-12}$. The long-term timing solution also allows us to obtain measurements of the proper motion of the pulsar: $\mu_{\alpha}\cos \delta = -5.56(9)$\,mas\,yr$^{-1}$ and $\mu_{\delta} = -2.2(6)$\,mas\,yr$^{-1}$, as well as a significant second spin frequency derivative $\ddot{f} = -2.733(7) \times 10^{-25}$\,Hz\,s$^{-2}$.

\subsection{Pulsars with MeerKAT timing solutions}

We obtained coherent timing solutions for three of the recent pulsar discoveries made with MeerKAT, namely M62G, M62H, and M62I, using the MeerKAT data spanning over three years. The timing residuals for these three pulsars as a function of MJD are shown in the bottom three panels of Figure \ref{fig:timing_mjd_plots}, and as a function of orbital phase in the corresponding panels of Figure \ref{fig:timing_orbphase_plots}. We discuss the results in the following subsections.

\subsubsection{M62G}

M62G is a pulsar in a binary system with an orbital period of $\sim18.6$\,hr and a $M_{\rm c, min}= 0.1$\,\msun. A partial-phase connection timing solution for this pulsar was previously reported in \citep{Ridolfi+2021}. To build the timing solution presented here, we extracted ToAs for every 20\,min from all of the MeerKAT observations. The ToAs were then fit to the ephemeris available from the partial-coherent solution. Since phase connection was not achieved at this point, we made use of the so-called ``jumps'' technique, introducing arbitrary phase jumps between the different observing epochs to refine the orbital parameters. We then removed as many arbitrary jumps as possible by estimating the exact number of phase rotations between the ToAs. We obtained a full-MeerKAT timing solution after a few iterations of the same procedure. Our measurements agree with the preliminary values given in \citet{Ridolfi+2021}, except that our measurement of the eccentricity $e=8.8(2)\times 10^{-5}$ is about one order of magnitude smaller. Additionally, measured the value of the longitude of periastron $\omega = 40.7(8)\deg$ for the first time. Our observations covered around 90\% of the orbit\footnote{The MeerKAT observations did not cover the orbital phases between 0.38 and 0.52 (see panel G of Figure \ref{fig:timing_orbphase_plots}).}, and we do not see any significant delays in the ToAs around the eclipse region; we therefore agree with the hypothesis that the companion is likely a white dwarf. 

\subsubsection{M62H}
\label{s:M62H}

To build the timing solution for M62H, we first obtained ToAs every 10\,min for all the observations from the orbital campaign. These were then fit with a pulsar model containing the spin frequency and orbital parameters ($P$, $x_{\rm p}$, $P_{\rm b}$, and $T_0$) derived using \texttt{fit\_circular\_orbit.py}, and the localisation from SeeKAT. Once we had the local timing solution from the orbital campaign, we then re-folded the MeerKAT data using this ephemeris.

This pulsar was detected with low S/N in the majority of the MeerTime observations, since those were pointed either towards the position of M62B or M62G. Consequently, we could only obtain a maximum of three ToAs per observation. Furthermore, the TRAPUM data proved to be sparse, preventing us from obtaining a coherent timing solution throughout the MeerKAT dataset in the first place. Therefore, we used the \DRACULA~code based on \texttt{TEMPO}\footnote{\url{https://tempo.sourceforge.net/}} \citep{Nice+2015} that is especially useful for
sparse data sets. This allowed us to determine a phase-connected timing solution for all the MeerKAT data (see Table \ref{tab:timing_fitresults_3}). The algorithm returned one possible solution with a reduced $\chi^2$ smaller than two after fitting for the position, spin period and
spin period derivative, and orbital parameters. We note that the position obtained with this solution is within the 1$\sigma$ error from \SeeKAT~(see Figure \ref{fig:M62_psrpositions}). The algorithm returned an orbital period of $P_{\rm b} = 0.132935028(8)$\,days and a  projected semi-major axis of $x_{\rm p}=0.004908(6)$\,lt-s. Assuming a pulsar mass of 1.4\,\msun, we obtain a minimum companion mass of only 0.00236\,\msun (2.5\,M$_{\rm J}$), indicating that the companion is one of the lightest known. There is no evidence for eclipses in any of the observations, suggesting that this pulsar can best be classified as a non-eclipsing black widow. 

\subsubsection{M62I}
\label{s:timing_M62I}

We followed a similar procedure as for M62H to build the timing solution for M62I. We obtained ToAs every 10\,min for the observations from the orbital campaign and fit them with a pulsar model derived using \texttt{fit\_circular\_orbit.py}. Once we had a timing solution from the orbital campaign, we then re-folded the MeerKAT data using this ephemeris. Following a similar procedure as for the timing of M62G, we obtained the phase-connected solution shown in Table 
\ref{tab:timing_fitresults_2}. Our measurements show that the eccentricity of this system ($e=2.08(9)\times 10^{-4}$) is small but non-zero. We also measure the longitude of periastron $\omega = 105(3)\deg$ and a loose constraint on the rate of change of the orbital period $\dot{P}_{\rm b} = -7(2)\times 10^{-12}$. M62I is located at $\alpha = 17\h01\m12\fs726(2)$ and $\delta = -30\degr06\arcmin50\farcs55(2)$, which places it 0.07\,arcmin south-east of the cluster centre as shown in Figure \ref{fig:M62_psrpositions}.

\subsection{Pulsars without full timing solutions}

\subsubsection{M62J}

We only have the two UHF detections of pulsar M62J and therefore we are currently unable to obtain a timing solution.

\subsection{Proper Motion}

The $\sim 23.5$-yr timing baseline provided by the Parkes and MeerKAT data enabled us to obtain measurements of the proper motion for all of the pulsars with long-term timing solutions (M62A-F) shown in Tables \ref{tab:timing_fitresults_1}-\ref{tab:timing_fitresults_3}. In Figure \ref{fig:M62_psrpositions}, we show the angular offsets of the pulsars in the sky with respect to the centre of the cluster. The direction and distance travelled by the pulsars in 1000\,yr are represented by the black arrows.  

The proper motion of the entire M62 cluster has been estimated by \citet{Vasiliev+Baumgardt2021} utilising data from the \textit{Gaia} Early Data Release 3, as $\mu_{\alpha} \cos \delta = -4.98(3)$\,mas\,yr$^{-1}$ and $\mu_{\delta} = -2.95(3)$\,mas\,yr$^{-1}$, for a total proper motion of 5.78(4)\,mas\,yr$^{-1}$. The proper motion of the cluster is shown as the black arrow in the bottom right corner of Figure \ref{fig:M62_psrpositions}. 

The $\mu_{\delta}$ measured from the timing solutions for most of the pulsars are around one sigma from the value of the GC. However, an exception is M62E, with $\mu_{\delta} = -1.9(4)$\,mas\,yr$^{-1}$, corresponding to a 2.6$\sigma$ deviation. In the case of the $\mu_{\alpha}$ values for pulsars M62B-D, the deviations remain within three sigma away from the value of the cluster. On the contrary, pulsars M62A, E and F exhibit significant deviations, at 4.3, 6.1, and 7.0 sigma away from the Gaia measurement. We note that the errors for $\mu_\delta$ are larger than the ones for $\mu_\alpha\cos\delta$. This is due to the fact that M62 is very close to the ecliptic plane.

We calculate the (unweighted) average proper motion from all pulsars in M62 as $\mu_{\alpha} \cos \delta = -5.0(2)$\,mas\,yr$^{-1}$, which is coincident with the GC value, and $\mu_{\delta} = -2.6(2)$\,mas\,yr$^{-1}$, indicating a 1.8$\sigma$ difference from the average value. We note that all of the pulsars are moving in the same general direction of the cluster, but apparently, the declination component is larger. 

Considering a distance of the cluster as in Section \ref{s:Intro}, the total proper motion implies that the pulsars are moving with transverse velocities in the range of $\sim 139$-171\,km\,s$^{-1}$. The pulsars' proper motions relative to the cluster suggest relative transverse velocities in the range of 2-26\,km\,s$^{-1}$ at the 1-$\sigma$ level. This is well within the estimated central escape velocity of the cluster of $\sim  59.3$\,km\,s$^{-1}$ \citep[value from Baumgardt's catalogue;][]{Baumgardt+Hilker2018}.

\subsection{Spin Period Derivatives}

\subsubsection{First spin period derivative}

Pulsars in globular clusters can be influenced by several acceleration effects, leading to modifications to the intrinsic (\Pdotint) to give the observed period derivative (\Pdotobs). We can use the following equation to account for these effects and determine (or constrain, depending on the case), the intrinsic period derivative \citep{Damour+Taylor1991}:

\begin{equation}
\label{eq:Pdot_obs}
\left(\dfrac{\dot{P}}{P}\right)_{\rm obs} = \left(\dfrac{\dot{P}}{P}\right)_{\rm int} + \dfrac{a_{\rm G} + a_{\rm GC} + a_{\rm PM}}{c} ,
\end{equation}

\noindent The different types of accelerations contributing to the observed period derivative are as follows: the difference between the accelerations of the Solar System and the cluster in the field of the Galaxy, denoted as $a_{\rm G}$; the line-of-sight component of the pulsar's acceleration caused by the gravitational field of the cluster, denoted as $a_{\rm GC}$; the centrifugal acceleration arising from the transverse Doppler effect, associated with the proper motion of the pulsar $\mu$ (where $a_{\rm PM} = \mu^2 D$ and D represents the distance between the GC and Earth) \citep{Shklovskii1970}.

We calculate the acceleration imparted by the Galaxy to M62, using equations 16 and 17 in \citealt{Lazaridis+2009}, the coordinates of M62, and using the values of the Solar motion in the Galaxy $\Theta_{0} = 240.5(4)$\,km\,s$^{-1}$, $R_{0} = 8.27(3)$\,kpc \citep{Guo+2021,Grav+2021}, as $a_{\rm G}$ $\approx$ 4.78 $\times$ $10^{-10}$\,m\,s$^{-2}$. The Shklovskii effect is computed as $\approx$ 1.38 $\times$ $10^{-10}$\,m\,s$^{-2}$ using the reported proper motion of M62 in \citet{Vasiliev+Baumgardt2021} and the same cluster distance as stated in Section \ref{s:Intro}. 
Since M62 is believed to be a core-collapsed or close to core-collapsed cluster, we cannot use analytical and numerical models to estimate the line-of-sight acceleration of the pulsar in the gravitational field, so we can obtain an estimate to within 10\% accuracy using the relation from \citet{Phinney1992, Phinney1993}, as follows:

\begin{equation}
\label{eq:max_a_GC}
    {\rm max}\frac{\abs{a_{\rm GC}}}{c} \simeq \frac{3}{2c}\frac{\sigma_v^2}{\sqrt{r_{\rm c}^2 + r_{\rm psr}^2}}
\end{equation}

\noindent for $r_{\rm psr} < 2 r_{\rm c}$, where $\sigma_v$ is the central velocity dispersion, $r_{\rm c}$ is the core radius, and $r_{\rm psr}$ is the projected distance of the pulsar from the centre of the cluster. We use $r_{\rm c} = 7\farcs9$, $\sigma_v = 14.8$\,km\,s$^{-1}$, and a distance $D=6.03$\,kpc from Baumgardt's catalogue.

We then calculate the maximum line of sight acceleration of the pulsars in terms of the observed period derivative $\dot{P}_{\rm obs}$ using the following equation:

\begin{equation}
\label{eq:al_max}
a_{l, \rm max} = \left(\dfrac{\dot{P}}{P}\right)_{\rm obs}c - \mu^2D - a_{\rm G}.
\end{equation}

\noindent The resulting values are displayed as black triangles pointing downwards in Figure \ref{fig:M62_accelerations} and represent an upper limit on the cluster acceleration.

The maximum line-of-sight acceleration due the cluster potential, $a_{l,\rm GC,max}$, is represented by the solid line in Figure \ref{fig:M62_accelerations} assuming the same model of equation \ref{eq:max_a_GC}. All of the pulsars have values of $a_{l,\rm max}$ that are smaller than the model $a_{l,\rm GC,max}$ for the line of sight. 

\subsubsection{Second spin frequency derivatives}

For most of the pulsars with long-term timing solutions in M62, the second spin frequency derivatives are significant. Yet, notably, most MSPs discovered within the Galactic disc exhibit undetectable timing noise. This leads to the hypothesis that the substantial second spin frequency derivatives observed in pulsars within GCs likely mirror their line-of-sight `jerk', denoted by $\dot{a}_l$, which  can be calculated using the following equation \citep{Joshi+Rasio1997}:

\begin{equation}
\label{eq:dot_al_max}
\frac{\dot{a}_{l}}{c} = \left(\dfrac{\dot{f}}{f}\right)^2 - \dfrac{\ddot{f}}{f}. 
\end{equation}

Our line-of-sight jerk measurement for the pulsars with significant $\ddot{f}$ values are reported in Tables \ref{tab:timing_fitresults_1}-\ref{tab:timing_fitresults_3}.
We can then estimate the maximum line-of-sight jerk induced by the motion of the pulsar in the field of the cluster $\dot{a}_{l,\rm GC,max}$ and the corresponding $\ddot{f}(\ddot{f}_{\rm max})$ in terms of the maximum velocity of the pulsar relative to the cluster, $v_{l,\rm max}$, using the following equation \citep{Freire+2017}:

\begin{equation}
\label{eq:dot_al_GC_max}
\dfrac{\dot{a}_{l, \rm GC, max}(0)}{c} = - \dfrac{\ddot{f}_{\rm max}}{f} = - \dfrac{4\pi G\rho(0)v_{l,\rm max}}{3c}. 
\end{equation}

\noindent As an estimate of $v_{l,\rm max}$, we use the same value of $\sigma_v$ and the core density $\rho(0)=8.5\times 10^{5}$\,\msun\,pc$^{-3}$ from Baumgardt's catalogue (see above). We then obtain $\dot{a}_{l,\rm GC,max}(0) = -2.38 \times 10^{-19}$\,m\,s$^{-2}$. 
All of the pulsars have line-of-sight jerks that are smaller than our estimate $\abs{\dot{a}_{l,\rm GC,max}}$ for their line of sight, we can therefore attribute the jerks measurements to the movement of the pulsars in the mean field of the cluster. 

\subsubsection{Third spin frequency derivatives}

The third and higher spin frequency derivatives can only be caused by the gravitational field of nearby stars \citep[see e.g.][]{Blandford+1987,Phinney1993,Joshi+Rasio1997,Freire+2017}. Pulsars M62B and M62C both have significant measurements of $f^{(3)}$ which could indicate the influence of nearby stars. 

\subsection{Orbital period derivatives}

We can also obtain an estimate of the line-of-sight acceleration from the gravitational field of the cluster in terms of the orbital period $P_{\rm b}$, and the observed orbital period derivative $\dot{P}_{\rm b,{\rm obs}}$ as follows:

\begin{equation}
\label{eq:a_GC}
a_{\rm GC} = \frac{\dot{P}_{\rm b,{\rm obs}}}{P_{\rm b}} c - \mu^2 D - a_{\rm G}.
\end{equation}

This is true in the case of pulsars where the intrinsic variation of the orbital period $\dot{P}_{\rm b,int}$ should be dominated by energy loss due to the emission of gravitational waves and is expected to be very small. However, this is unlikely to be the case for the eclipsing systems, where the presence of higher orbital frequency derivatives allows a description of the orbital behaviour, but  the model should be seen as a display of the unpredictable orbital variability \citep[see e.g.][]{Freire+2017, Ridolfi+2016}.

\begin{figure}
\centering
	\includegraphics[width=\columnwidth]{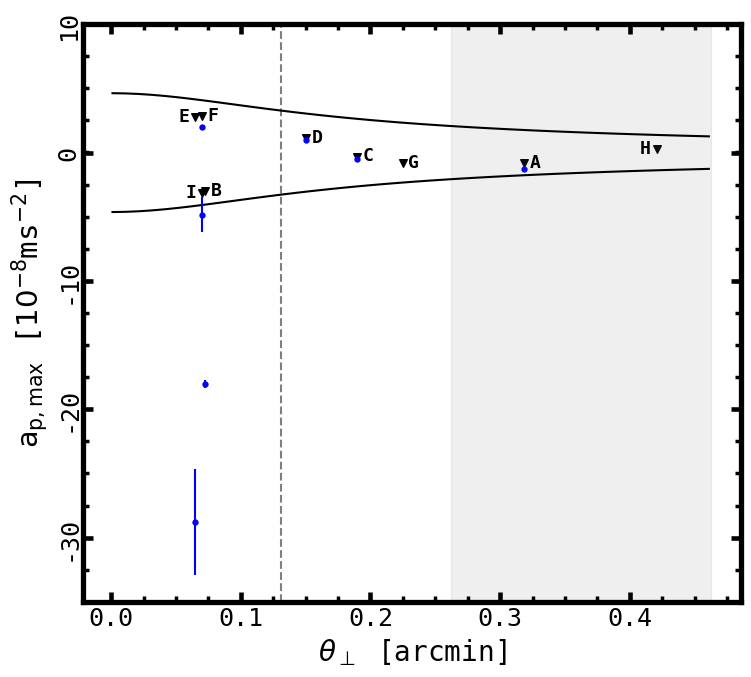}
    \caption{Line-of-sight acceleration as a function of the total angular offset from the centre of the cluster ($\theta_{\bot}$) for pulsars with timing solutions in M62. Upper limits for the accelerations of the MSPs are shown as the black triangles, determined from $\dot{P}_{\rm obs}$. The measurements of the line-of-sight accelerations for pulsars determined from $\dot{P}_{\rm b,obs}$ (where available) are shown as the blue points and error bars. For most of the pulsars these errors are very small as determined by their timing solutions. The maximum and minimum acceleration ($a_{l,{\rm GC, max}}$) along the line of sight predicted by using Equation \ref{eq:max_a_GC} are shown as the black solid lines.
    The core radius is indicated by the vertical dashed line. The shaded area highlights the region >2$r_{\rm c}$ where the approximation of the same equation is only valid to 50\%.}
        \label{fig:M62_accelerations}
\end{figure}

The accelerations measured in terms of $\dot{P}_{\rm b,{\rm obs}}$  are presented in Tables \ref{tab:timing_fitresults_1}-\ref{tab:timing_fitresults_3}, and are also shown as the vertical blue error bars in Figure \ref{fig:M62_accelerations}. All these accelerations have similar values as the measured $a_{l,\rm max}$ (see Figure \ref{fig:M62_accelerations}), except for the case of the two eclipsing pulsars. We attribute this to the unpredictable variations in the orbital period with time as seen for other eclipsing systems \citep[see e.g.][]{Prager+2017, Freire+2017}.

\subsection{Intrinsic spin period derivatives}

We notice that the measured values of $a_{l, \rm GC}$ (for the non-eclipsing binaries) tend to be slightly smaller than $a_{\rm max}$. The difference is likely due to the contribution from $\dot{P}_{\rm int}$, we can then derive $\dot{P}_{\rm int}$ by combining equations \ref{eq:Pdot_obs}
 and \ref{eq:a_GC}, considering that the $\dot{P}_{\rm b,int}$ is small:
 
\begin{equation}
\label{eq:Pdot}
\dot{P}_{\rm int} = \dot{P}_{\rm obs} - \frac{\dot{P}_{\rm b,{\rm obs}}}{P_{\rm b}}P.
\end{equation}

\noindent The resultant $\dot{P}_{\rm int}$ are presented in Tables \ref{tab:timing_fitresults_1} and \ref{tab:timing_fitresults_2}. We then explicitly calculate the characteristic ages, $\tau_c = P/(2\dot{P}_{\rm int})$, and the surface magnetic fields $B_0 = 3.2 \times 10^{19} \sqrt{P \dot{P}}$. These are also presented in Tables \ref{tab:timing_fitresults_1}-\ref{tab:timing_fitresults_3}.

\subsection{Possible X-ray counterparts and location in the cluster}

Figure \ref{fig:M62_psrpositions} shows the positions of all pulsars (as black dots) and the X-ray sources from \citet[][purple dots]{Oh+2020} known in the cluster relative to its centre. Using our best positions we find that none of the pulsars are clearly coincident with any of the X-ray sources. 

\citet{Cocozza+2008} claimed the association of M62B with an X-ray source located at $\alpha = 17\h01\m12\fs70$, $\delta = -30\degr06\arcmin49\farcs08$ with an error box of 0.7\,arcsec; this position is 0.42\,arcsec away from our best position of M62B. However, \citet{Oh+2020} added more Chandra data and obtained a position of the source which places it 0.02\,arcsec further away from our measurement, with an error of 0.056\,arcsec in right ascension and 0.051\,arcsec in declination. We then find a positional offset between M62B and the s22 Xray source of 0.463(2)\,arcsec. Due to the presence of several spin frequency derivatives and orbital frequency derivatives, there might be a covariance with the absolute timing position. In order to verify this non-association, we re-fitted our timing solution using the position of the s22 Xray source. However, there is a clear sinusoidal shape in our residuals typical of a position error, supporting the non-association. \citet{Oh+2020} also suggested that M62C was coincident with the X-ray source s10. We find the offset between M62C and source s10 is 0.814(2)\,arcsec, and so we find an association to be unlikely. 

In Table \ref{tab:pulsar_offsets}, we report the angular offsets of all pulsar positions in the sky relative to the centre of the GC. The projected distances ($r_\bot$) were calculated using the most recent GC parameters from Baumgardt's catalogue, and are also reported in Table \ref{tab:pulsar_offsets}. Pulsars M62B, E, F, and I are found within the GC core (0.13\,arcmin), while pulsars M62A, C, G, H, and J are outside the core but within the half-light radius. Moreover, all of the pulsars are well within the distance of M62H to the centre of the cluster of 0.4\,arcmin. This distance corresponds to approximately 27\% of the cluster's half-mass radius, measuring 1.66\,arcmin. The observed pulsar distribution is similar to the dynamically relaxed pulsar population in GCs like 47 Tuc,  which goes to about 3 core radii. Unlike what is observed in most core-collapsed GCs \citep{Verbunt+Freire2014}, there are no pulsars farther out that show signs of recent recoils from exchange encounters. We searched MeerKAT beams out to a radius of 3\,arcmin, so this is not a selection effect.

\begin{table}
\setlength\tabcolsep{7pt}
  \begin{threeparttable}
    \caption{Pulsar offsets from the centre of M62.}
     \label{tab:pulsar_offsets}
    \renewcommand{\arraystretch}{1.0}
     \begin{tabular}{lccccc}
     \hline
      \hline
 & & & \multicolumn{2}{c}{$\theta_{\bot}$} & \\
 & $\theta_{\alpha}$\tnote{*} & $\theta_{\delta}$\tnote{*} &  \multicolumn{2}{c}{\rule{2cm}{0.2pt}} & $r_{\bot}$   \\
Pulsar & (arcmin) & (arcmin) & (arcmin) &  ($\theta_{\rm c}$) & (pc)  \\        
        \midrule
A    &  -0.1053  & 0.3006 & 0.32  &  2.43  &  0.56 \\
B    &  -0.0710  & -0.0139  & 0.07  &  0.55  &  0.13 \\
C    &  -0.0285  &  -0.1875 &  0.19  &  1.45  &  0.33 \\ 
D    &  0.1211  & 0.0888  &  0.15  &  1.16  &   0.26  \\
E    &  0.0604  &  0.0223  &  0.06  & 0.49  &  0.11  \\
F    &  -0.0376  &  -0.0588  &  0.07  &  0.53  &  0.12  \\
G    &  0.1932 & 0.1153  &  0.23  &  1.72  &  0.39  \\
H    &  0.1713  & 0.3843  &  0.42   &  3.21  &  0.74  \\  
I    &  -0.0585  & -0.0390  &  0.07  &  0.54  &  0.12  \\  
J    &  0.1827  & -0.0399  &  0.06  &  0.47  &  0.11  \\  
    \hline
      \hline
     \end{tabular}
    \begin{tablenotes}
      \small
      \item[*] The uncertainties are much smaller than the uncertainty of the GC's centre, assumed to be exactly as indicated in Figure \ref{fig:M62_psrpositions}, except for the case of M62J. The 1-$\sigma$ errors for this pulsar are $\theta_{\alpha} = 0.3$\,arcmin and $\theta_{\delta} = 0.3$\,arcmin.
    \end{tablenotes}
  \end{threeparttable}
\end{table}

\begin{figure*}
	\includegraphics[width=\textwidth]{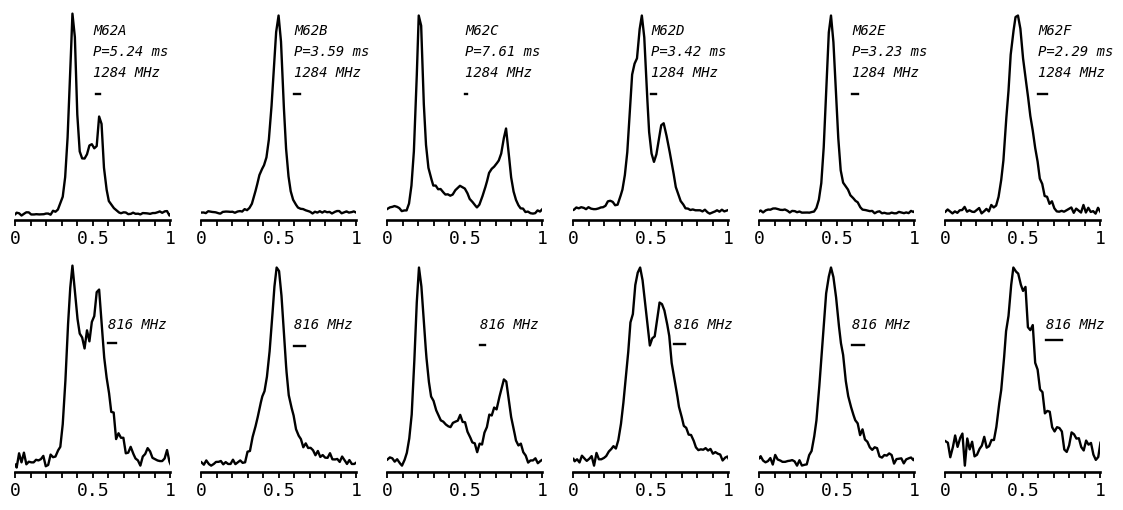}
    \caption{Integrated pulse profiles for all of the previously known pulsars using both L-band (1284 MHz) and UHF (816 MHz) data from TRAPUM. All profiles show one full rotation of the pulsar with 64 phase bins. Their spin periods are indicated. The length of the horizontal bar indicates the effective time resolution of the system, relative to each pulsar’s spin period.} 
    \label{fig:integrated_profiles_known}
\end{figure*}

\begin{figure}
\centering
	\includegraphics[width=\columnwidth]{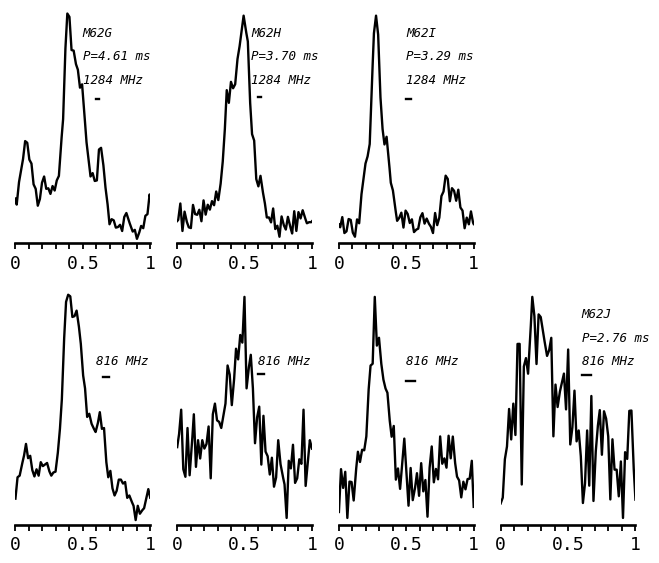}
    \caption{Integrated pulse profiles for recent discovered pulsars using both L-band (1284\,MHz, except for M62J) and UHF (816\,MHz) data from TRAPUM. All profiles show one full rotation of the pulsar with 64 phase bins. Their spin periods are indicated. The length of the horizontal bar indicates the effective time resolution of the system, relative to each pulsar’s spin period.}
    \label{fig:Pulse_profiles_new} 
\end{figure}

\subsection{Flux Densities}

The flux densities at $\sim 1284$\,MHz ($S_{\rm 1284}$) of all the pulsars and their pseudo-luminosities ($L_{1284} \equiv S_{1284}D^2$) are calculated using the radiometer equation \citep{Dewey+1985} and using the same cluster distance as established in Section \ref{s:Intro}. To calculate the system equivalent flux density, we then use a system temperature $T_{\rm sys}$ = 31.3\,K, which includes the sky temperature at the cluster centre, estimated using the \texttt{Python} implementation of the Global Sky Model (GSM2008 of \citet{dOliveira-Costa+2008} from the radio sky \texttt{PyGSM}\footnote{\url{https://github.com/telegraphic/PyGSM}}, $T_{\rm sky} = 8.8$\,K; the atmosphere plus the ground spillover temperature $T_{\rm atm+spill}$ ($\sim 4.5$\,K at 45\degr elevation\footnote{{ \url{https://skaafrica.atlassian.net/rest/servicedesk/knowledgebase/latest/articles/view/277315585} }}); the receiver temperature $T_{\rm rec} = 18$\,K; and the gain of the telescope $G = 2.62$\,K\,Jy$^{-1}$ (for the MeerKAT array observations using 60 antennas). To account for the various sensitivity losses due to signal processing and digitisation, we assumed a correction factor of 1.1. The flux densities and the pseudo-luminosities were determined using the widths and S/N values of the integrated pulse profiles from the top panels of Figure \ref{fig:integrated_profiles_known} for the case of the previously known pulsars and of Figure \ref{fig:Pulse_profiles_new} for the recently discovered pulsars. The profiles were obtained by summing together (without including weights) all the TRAPUM data from the beams that pointed directly at the positions of the pulsars, using \texttt{prsadd} from the \texttt{PSRCHIVE} package, giving a total of 22.4\,hr. We excluded all subintegrations in time or frequency that are corrupted by significant RFI in order to increase the S/N. For the eclipsing pulsars (M62B and M62E), we also excluded the subintegrations where the signal of the pulsar is eclipsed.  

For the case of M62H and M62I,  the profiles were obtained by combining all TRAPUM observations to increase the S/N. However, the S/N and the pulse width values used to calculate $S_{1284}$ and $L_{1284}$ were obtained using only the observations from the orbital campaign, summing together a total of 10\,hr. This was done in order to have a more realistic measurement of the S/N, since for the other MeerKAT observations, none of the beams were directly pointing towards their positions. We could not obtain an L-band profile for M62J, since we do not have any detections at this frequency. Nevertheless, upper limits for its flux density and pseudo-luminosity were obtained, assuming a S/N of 3 and the same pulse width as measured at UHF (see below). 

For comparison, the 816\,MHz flux densities and their pseudo-luminosities were also obtained following a similar procedure, but considering a $T_{\rm sys}$ = 46.5\,K, which includes the sky temperature at the cluster centre estimated using \texttt{PyGSM} (see above), $T_{\rm sky} = 20$\,K; the atmosphere plus the ground spillover temperature $T_{\rm atm+spill}$ ($\sim 6.5$\,K at 45\degr elevation\footnote{{\url{https://skaafrica.atlassian.net/rest/servicedesk/knowledgebase/latest/articles/view/277315585}}}); the receiver temperature $T_{\rm rec}$ (18\,K); and the gain of the telescope $G = 2.62$\,K\,Jy$^{-1}$ (for the MeerKAT array observations using 60 antennas), and using the UHF TRAPUM data of the 2 observations summing a total of 3\,hr (without including weights) from the beams that pointed directly to the position of all the pulsars, except for M62F and M62H-J. For M62F, the data from the beam of the first observation pointing to the pulsar's position was corrupted. Therefore, the UHF profile includes only the 2\,hr of the follow-up UHF observation (see Table \ref{t:obs_summary}). For pulsars M62H-J we used the data from the beams that were pointed closest to their position. M62J was detected only in the full hour of the 1st UHF observation and in a 30\,min segment of the follow-up UHF observation, giving a total of 90\,min. The average UHF pulse profiles are shown in the bottom panels of Figures \ref{fig:integrated_profiles_known} and \ref{fig:Pulse_profiles_new}. The profiles at the two different frequencies were aligned by using a cross-correlation method. 

The results for the pulsars with timing solutions are shown in Tables \ref{tab:timing_fitresults_1}-\ref{tab:timing_fitresults_3}. For M62J, we estimate the mean flux density at $S_{\rm 816} = 0.08$\,mJy and $L_{816} = 5.19$\,mJy\,kpc$^2$. At L-band, we calculated the upper limits of the mean flux density and the pseudo-luminosity using a S/N=3 and the same pulse width as in UHF to be $S_{\rm 1284} < 0.02$\,mJy and the pseudo-luminosity $L_{1284} < 1.28$\,mJy\,kpc$^2$, obtaining a spectral index $\beta < -3.09$.

\subsubsection{Profile Widths and Shapes}
 
To measure the observed pulse width at 10\% and 50\% of the peak intensity (referred to as W10 and W50, respectively), we again used the template we formed to perform the timing analysis. We utilised the concentration parameter obtained from the best-fit combination of von Mises functions and the \texttt{fitvonMises} function from the \texttt{PSRSALSA} software \citep{Weltevrede2016} for this purpose. To quantify the error associated with the width measurement, we simulated 1000 realisations of each profile using the off-pulse noise statistics derived from the average profile. Subsequently, we repeated the von Mises fitting process for each simulated profile and determined the pulse width in each case. The error on the width measurement was estimated from the distribution of width values obtained from these simulations. The resulting widths and errors are shown in Table \ref{tab:pulsar_widths}. We describe below the profile shapes and the frequency dependence of each of the pulsars to be used for future consideration of MSP pulse profile shapes.

Pulsar M62A has a complex pulse profile characterised by three distinct peaks across its main pulse. These peaks are visible in both L-band and UHF frequencies. Notably, the second and third peaks are more prominent at UHF frequencies, indicating profile evolution. The UHF profile is $\sim 3$ times wider than at L-band at 50\% of pulse maximum, due to the increase of the trailing component. The profile of M62B, on the other hand, is composed by one peak but has a broad feature preceding the main pulse, which is somewhat more prominent at L-band than at UHF. At 50\% of the pulse maximum, the UHF profile is $\sim 1.3$ times wider than the L-band profile. This pulsar is the brightest at L-band, with a pseudo-luminosity $L_{1284} = 17.70$\,mJy\,kpc$^2$. 

\begin{table}
\setlength\tabcolsep{7pt}
  \begin{threeparttable}
    \caption{Pulse widths at 50\% (W50) and 10\% (W10) measured at L-band and UHF in degrees of rotation by fitting von Mises functions to the average profiles which are then used to estimate the pulse duty cycle W50/P. The profile can either be fitted with a single (S), or multiple (M) component model.}
     \label{tab:pulsar_widths}
    \renewcommand{\arraystretch}{1.0}
     \begin{tabular}{lccccc}
     \hline
      \hline
Pulsar & No. & Freq &  W50 & W10 & W50/P  \\
 & Comp & (MHz) & (deg) &  (deg) & (\%)  \\        
        \midrule
A    &  M  & 1300 &  29(21) & 100.3(7)  &  8.1 \\
     &  M  & 816 & 88.8(8)  &  135(2)   &  24.7\\
B    &  M  & 1300 & 30.3(1)  &  83.3(4)  &  8.4 \\
     &  M  & 816 & 40.2(5)  &  114(2)  &  11.2 \\
C    &  M  & 1300 &  20.3(4)  &  168(6)  &  5.7 \\ 
     &  M  & 816 &  30.6(4)  &  225(6)  &  8.5\\ 
D    &  M  & 1300 & 42.3(2)  &  125.8(6)  &  11.8 \\
     &  M  & 816 &  103.8(6)  &  168(2)  &  28.8 \\
E    &  M  & 1300 &  27.1(1)  &  67(1)  &  7.5 \\
     &  M  & 816 &  52.7(5)  &  127(2)  &  14.7 \\
F    &  S  & 1300 &  53.3(5)  &  103.1(9)  &  14.8 \\ 
     &  M  & 816 &  73(2)  &  127(3)  &  20.3 \\ 
G    &  M  & 1300 & 54(2) & 241(3) & 15.2 \\
     &  M  & 816 & 63(2)  & 247(5) & 17.5 \\
H    &  M  & 1300 & 66(2) & 122(8) & 18.4 \\
     &  M  & 816 & 72(11) & 153(14) & 20.06 \\   
I    &  M  & 1300 & 31(2) & 105(44) & 8.2 \\
     &  M  & 816 & 43(5) & 104(16) & 12.0 \\
J    &  M  & 816 & 122(8) & 172(9) & 33.84 \\     
    \hline
      \hline
     \end{tabular}
  \end{threeparttable}
\end{table}

The profile of pulsar M62C has four clear peaks across the main pulse, with the first peak being the most prominent, followed by the fourth, third, and second peaks in consecutive order. These peaks are visible in the UHF profile as well, although the prominence of the third peak diminishes, and the second peak becomes more prominent at lower frequencies. The width at 50\% of pulse maximum in the UHF profile is $\sim 1.5$ times that of the L-band. Pulsar M62D showcases a similarly intricate profile with two main peaks, with the first being the most prominent of the two. TRAPUM observations reveal a broad feature preceding the main pulse. Notably, the second peak is more prominent at UHF, and the profile is $\sim 2.5$ times wider at 50\% of pulse maximum at 816\,MHz. This pulsar is the brightest of the pulsars at UHF ($L_{816}$ = 52.83\,mJy\,kpc$^{2}$) and is the pulsar with the steepest spectral index $\beta = -4.04$.

The profile of M62E can be described as a single peak with what could be assumed to be a scattering tail, observable at both L-band and UHF frequencies, longer at lower frequency, as expected from scattering since it is strongly frequency dependent (i.e. $\propto\nu^{-4}$). Apparent scattering tails can be seen in some of the pulsars, but this effect is more evident for pulsar M62E. At 50\% of the pulse maximum, this profile is approximately 1.9 times wider in UHF than in L-band, which is lower than the expected value from scattering. We then use the PyGEDM: Galactic Electron Density Models\footnote{\url{https://apps.datacentral.org.au/pygedm/}} to estimate the scattering timescales. The YMW16 model \citep{YM2016} predicts the scattering timescale at 1\,GHz  and at the highest DM (DM value of pulsar M62A, 114.98\,pc\,cm$^{-3}$) could be as much as 83\,$\mu$s, whereas the NE2001 model \citep{NE2001} estimates (also at 1\,GHz) this could be as much as 18\,$\mu$s at UHF, which represents only 0.4-2\% of the average spin period of pulsars in M62 and therefore scattering is unlikely to be affecting these profiles according to these models. However, these scattering estimates are very uncertain since $b = 7.317\degr$. The fact that not only M62E but M62A, M62B, M62D and M62F also show exponential tails in their profiles can be an indication of the presence of scattering. However, due to the dispersion smearing present in this data set, we cannot constrain this well and leave it for a more detailed future study. The profile of pulsar M62F displays some unresolved features in the L-band profile, whereas the profile in UHF is noisier, since it was obtained using only 2\,hr from the 2nd UHF observation (see Table \ref{t:obs_summary}). The UHF profile at 50\% of pulse maximum is $\sim 1.4$ the width at L-band. 

M62G is the pulsar with the most complex profile in M62, showing five peaks in the main profile at L-band frequencies, with the central peak being the most prominent. The same peaks are observed at UHF, with an additional peak noticeable after the most prominent one. The profile  then shows evolution, and the width at 50\% of the pulse maximum is approximately 1.2 times that of L-band. The L-band profile of pulsar M62H consists of a central peak, preceded by two peaks and followed by one. The W50 at this frequency is approximately the same as at UHF. However, conclusive details can not be determined due to the low S/N of the profile at UHF. The L-band profile of M62I displays a main pulse at around phase 0.4 and an inter-pulse at around phase 0.7. The main component also shows two broad features both preceding and following the main pulse. The profile at UHF is noisier but all these components are still visible. The W50 at UHF is approximately 1.4 times that of the L-band. Finally, the profile of M62J is not well resolved, due to the very low S/N. We estimate a width at 50\% of pulse maximum at UHF of 121(7)\,$\deg$.

In summary, we note that all profiles are broader at 50\% of pulse maximum at UHF than at L-band ranging from $\sim 1$ (for the case of M62H), up to a maximum factor of $\sim 3$ for M62A. As can be seen in the horizontal bars indicating the effective time resolution in Figures \ref{fig:integrated_profiles_known} and \ref{fig:Pulse_profiles_new}, this is not an instrumental effect. 

\subsection{Eclipses of M62B and M62E}

Pulsars M62B and M62E are black widow and redback systems, respectively, and they exhibit eclipses. Here we use the sensitive, wide-band and multi-frequency MeerKAT data to study their eclipses. Our data set consists of 23 epochs at the central frequency of 1284\,MHz, and two epochs at the central frequency of 816\,MHz. The two frequency bands were subdivided into two subbands to study the frequency dependencies of the eclipsing behaviour. We discuss the results for both pulsars in the following subsections. 

\subsubsection{M62B}

\begin{figure}
\centering
	\includegraphics[width=\columnwidth]{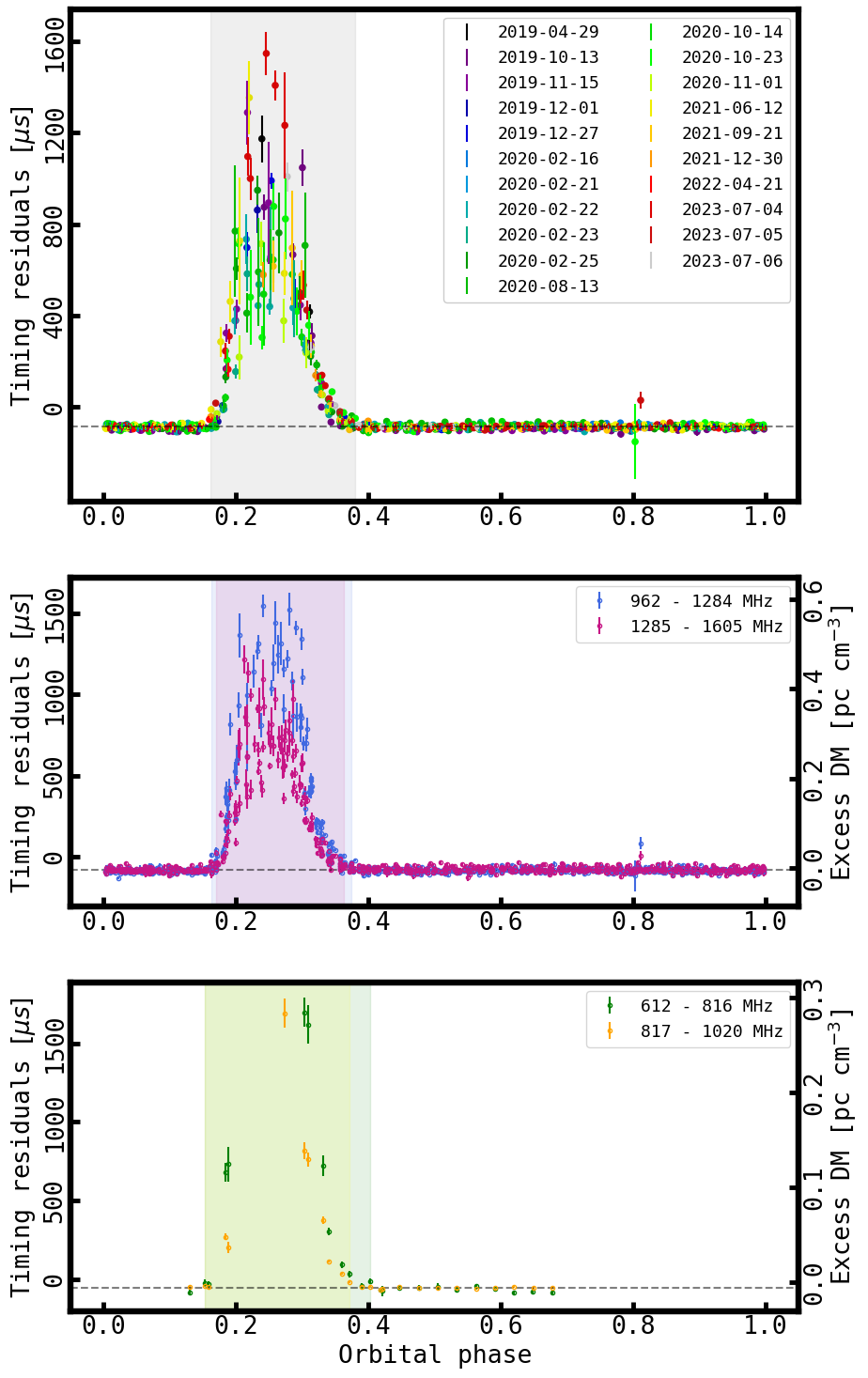}
    \caption{Top panel: Timing residuals as a function of orbital phase for the eclipsing binary pulsar M62B. The different colours indicate the different epochs at 962--1605\,MHz. Middle panel: Timing residuals and excess DM measured simultaneously at 962--1284\,MHz (blue) and 1285--1605\,MHz (pink). The eclipse regions at both subbands are highlighted covering the respective orbital phases.
    Bottom panel: Timing residuals and excess DM measured simultaneously at
    612--816\,MHz (green) and 817--1020\,MHz (yellow). The eclipse regions for the different subbands are highlighted in the same colours by covering the respective orbital phases.}
        \label{fig:M62B_eclipses}
\end{figure}

\textbf{L-band}

The timing residuals shown in Figure \ref{fig:M62B_eclipses} were generated using an ephemeris based on the out-of-eclipse ToAs using \TEMPOTWO~(see Table \ref{tab:timing_fitresults_3}) for the 23 observing epochs (denoted by different colours) as a function of orbital phase. These are presented in the top panel of Figure \ref{fig:M62B_eclipses}. The specific regions of the eclipse that these observations covered are listed in Table \ref{t:obs_summary}. Delays of $\sim 1$\,ms in the times of arrival can be seen before, after and through the eclipse. The eclipse region is defined by those ToAs which deviate by more than 3$\sigma$ above the mean value of the out-of-eclipse ToAs. The maximum delay during the eclipse is $\sim 1.5$\,ms at a phase of $\sim 0.24$. Ingress and egress transitions are spread over a range of orbital phases, with the ingress starting at orbital phase $\sim 0.16$ and egress ending at $\sim 0.38$, giving a total span of the eclipse of approximately 22\% of the orbit.

Using the mass function from our timing solution, $f(M_{\rm p},M_{\rm c}) = 8.29849(8)$ $\times$ 10$^{-4}$\,\msun, assuming a pulsar mass $M_{\rm p} = 1.4$\,\msun\, and an inclination angle $i = 60^{\circ}$, we then obtain a companion mass $M_{\rm c} \sim 0.13$\,\msun. The separation of the binary components corresponds to an eclipsing region with physical size of R$_{\rm E}$ $\sim 0.87$\,\rsun, which is larger than the Roche Lobe radius of the companion R$_{\rm L} \simeq 0.34$\,\rsun, indicating that the eclipsing material is not gravitationally bound to the companion and that the companion is losing mass \citep{Stappers+1996}. This might be supporting evidence for the claim by \citet{Cocozza+2008} of a tidally deformed companion star which is losing mass.

In order to probe the frequency dependence of the eclipse characteristics of this pulsar, we divided the effective 643\,MHz bandwith (after RFI cleaning) into two subbands, obtaining ToAs every 3\,min. The eclipse regions for both subbands are shown as coloured regions in the middle panel of Figure \ref{fig:M62B_eclipses}. The maximum delay in timing residuals around eclipse transition is $\sim 1.5$\,ms at orbital phase $\sim 0.24$, while at higher frequencies the maximum delay is $\sim 1.2$\,ms at $\sim 0.21$ in phase. The eclipse ingress and egress transitions exhibit a potential asymmetry in frequency dependence. In the 962-1284 MHz band, the ingress begins earlier at $\sim 0.16$ in phase, while the egress ends later at $\sim 0.38$, resulting in a total eclipse span of approximately 22\% of the orbit. Whereas in the 1282--1605\,MHz band, the ingress starts at $\sim 0.17$ and egress ends at $\sim 0.36$, giving a total eclipse span of 19\% of the orbit. The eclipse duration is then $\sim 1.16$ times longer for the 962--1284\,MHz band compared to the 1282--1605\,MHz band. Interestingly, we find that ToAs from the observation on 2023-07-05-16:27 displayed delays away from the eclipse region in both subbands, in the lower band of 8\,$\mu$s, and in the higher band of 6\,$\mu$s, hinting at material in the orbit, at orbital phase of $\sim 0.81$. 

The variation in the excess DM with orbital phase is also shown in Figure \ref{fig:M62B_eclipses}. The presence of an additional electron column density in the eclipse region implied by this dispersive delay, that is, the maximum added electron density near superior conjunction, is found to be $N_{e,{\rm max}} \geq$ 1.9 $\times$ 10$^{15}$\,cm$^{-2}$. We estimate the corresponding electron density in the eclipse region ($n_e \sim N_e/2a$, where $a = a_{\rm p} + a_{\rm c} \approx 1.3$\,\rsun, assuming an inclination $i = 60^{\circ} $) as 1.04 $\times$ 10$^{4}$\,cm$^{-3}$. 

\textbf{UHF}

We further investigate the frequency dependence of the eclipse by splitting the effective bandwith of the two UHF observations (see Table \ref{t:obs_summary}), which covers only around 55\% of the orbit, into two bands: 612--816\,MHz and 817--1020\,MHz. We also obtained one ToA every 3\,min in this case. The results are shown in the bottom panel of Figure \ref{fig:M62B_eclipses}. The maximum delay in the timing residuals around eclipse transitions is $\sim 1.7$\,ms at $\sim 0.30$ in phase for the lower UHF frequencies (612--816\,MHz, green colour). At the higher UHF frequencies (817--1020\,MHz, yellow colour) the maximum delay around eclipse transition is $\sim 1.7$\,ms at $\sim 0.27$ in phase. The variation in the excess DM with orbital phase is also shown in the same figure. The different eclipse regions are shown as coloured areas in the bottom panel of Figure \ref{fig:M62B_eclipses}. At lower UHF frequencies the ingress transition starts at orbital phase $\sim 0.15$, while the eclipse egress ends at $\sim 0.40$, giving a total span of the eclipse of approximately a quarter (25\%) of the orbit. At higher UHF frequencies, the ingress starts at orbital phase $\sim 0.15$, while the egress ends at $\sim 0.37$ giving a total span of the eclipse of $\sim 22$\% of the orbit. We observe that the eclipse duration is $\sim 1.14$ times longer for the 612--816\,MHz band than for the 817--1020\,MHz band. 

\subsubsection{M62E}

\begin{figure}
\centering
	\includegraphics[width=\columnwidth]{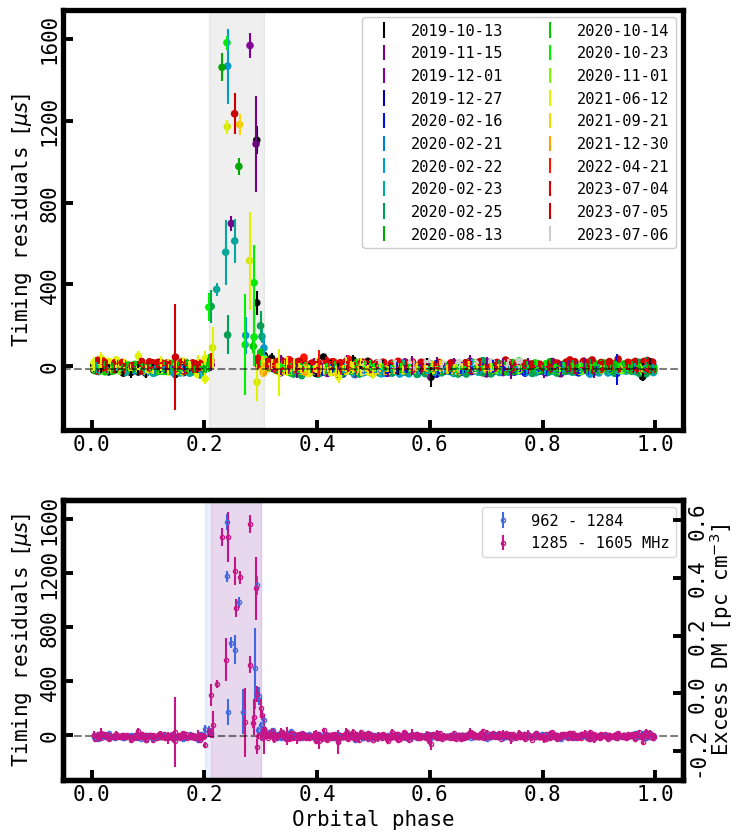}
    \caption{Top panel: Timing residuals as a function of orbital phase for the eclipsing binary pulsar M62E. The different colours indicate the different epochs at 962--1605\,MHz. Bottom panel: Timing residuals and excess DM measured simultaneously at 9621--1284\,MHz (blue) and 1285--1605\,MHz (pink). The eclipse regions at both subbands are highlighted in the same colours by covering the respective orbital phases.}
        \label{fig:M62E_eclipses}
\end{figure}

\textbf{L-band}

We study the L-band eclipses for M62E using a similar method as that followed for M62B. Timing residuals shown in Figure \ref{fig:M62E_eclipses} were generated using an ephemeris based on our out-of-eclipse ToAs using \TEMPOTWO~from the 23 observing epochs (denoted by different colours, see Table \ref{tab:timing_fitresults_3}) as a function of orbital phase. The specific regions of the eclipse that these observations covered are listed in Table \ref{t:obs_summary}. Delays of $\sim 1$\,ms in the times of arrival can be seen before, after, and sometimes through the eclipse. The maximum delay in the timing residuals during the eclipse is $\sim 1.6$\,ms at $\sim 0.24$ in phase as visible in the top panel of Figure \ref{fig:M62E_eclipses}. The eclipse ingress and egress transitions are spread over a range of orbital phases. The ingress transition starts at orbital phase $\sim 0.21$, while the eclipse egress ends at orbital phase $\sim 0.31$, giving a total span of the eclipse of approximately 10\% of the orbit. 

Using the mass function from timing, $f(M_{\rm p},M_{\rm c}) = 1.47578(3) \times 10^{-5}$ \,\msun, and assuming a pulsar mass $M_{\rm p} = 1.4$\,\msun\ and an inclination angle $i = 60^{\circ}$, we calculate a companion mass $M_{\rm c} = 0.03$\,\msun. The separation of the binary components corresponds to an eclipsing region with physical size of R$_{\rm E}$ $\sim 0.43$\,\rsun, which is almost twice the size of the Roche lobe radius of the companion R$_{\rm L} \simeq 0.22$\,\rsun, indicating that the eclipsing material is not gravitationally bound to the companion and that the companion is losing mass \citep{Stappers+1996}. We also calculate the energy density of the pulsar wind at the companion distance, $U_{\rm E} = \dot{E} / 4\pi^2 ca^2$, and obtain $U_{\rm E} = $32.46\,erg\,cm$^{-3}$ using the $\dot{P}$ from our timing solution. 

We also study the frequency dependence of the eclipse for this pulsar by dividing the observed bandwidth into two sub-bands for the L-band observations as seen in the bottom panel of Figure \ref{fig:M62E_eclipses}. The maximum delay in timing residuals around eclipse transitions is $\sim 1.6$\,ms at $\sim 0.24$ in phase for the lower frequencies (962--1283\,MHz, blue colour in Figure \ref{fig:M62E_eclipses}). At higher frequencies (1284--1605 MHz, pink colour in Figure \ref{fig:M62E_eclipses}) the maximum delay around eclipse transitions is $\sim 1.5$\,ms at $\sim 0.28$ in phase. We observe a larger eclipse duration at the lower frequency band ($\sim 1.1$ times longer for the 962--1283\,MHz band than in the 1284--1605\,MHz band), showing a possible asymmetry in the frequency dependence of eclipses between the two subbands. Different ingress/egress durations that depend on the frequency have also been observed for other spider pulsars \citep[e.g.][]{Kudale+2020,Polzin+2020}. 

The variation in the excess DM with orbital phase is also shown in the bottom panel of Figure \ref{fig:M62E_eclipses}. The presence of an additional electron column density in the eclipse region implied by this dispersive delay, that is, the maximum added electron density near superior conjunction, is found to be $N_{e,{\rm max}} \geq$ 2.57 $\times$ 10$^{15}$\,cm$^{-2}$. We estimate the corresponding electron density in the eclipse region ($n_e \sim N_e/2a$, where $a = a_{\rm p} + a_{\rm c} \approx 1.4$\,\rsun, assuming an inclination $i = 60^{\circ} $) as 1.32 $\times$ 10$^{4}$\,cm$^{-3}$.

For the case of the UHF observations, only observation 2022-04-21-01:55 covered the eclipse region. This observation suggests DM variations of $8\times 10^{-3}$\,pc\,cm$^{-3}$. 

\section{Discussion}
\label{s:Discussion}

In this paper, we report the discovery of the MSPs M62H, M62I and M62J in the GC M62. The three new discoveries bring the number of pulsars in this cluster to ten, all of them in binary systems. We note that for a given rotation period and flux density value, isolated pulsars are easier to find than binaries. This is because the spin period of isolated pulsars does not undergo periodic variations due to the orbital motion. The fact that none have been seen so far is therefore not an observational selection effect. We also present the updated long-term timing solutions, with a data span of over 23\,yr, for the previously known pulsars (M62A-F), along with the MeerKAT-based timing solutions of M62G, M62H, and M62I, which are presented here for the first time. 
The large timing baselines allow us to measure pulsar parameters like proper motions, orbital period derivatives and jerks, which in turn allow us to determine the line-of-sight accelerations. All these results will be useful for future studies that can investigate the mass distribution of the cluster and perhaps confirm or rule out the evidence that has been found for non-luminous matter in the centre of M62 \citep{Abbate+2019}. Of the three new pulsars discovered, M62I is the only new addition that can contribute to a measurement of the mass distribution, due to its distance from the cluster core of 0.7\,arcmin, similar to values of pulsars M62B, M62E and M62F, which were used in \citet{Abbate+2019}. It is important to note that the values of the cluster parameters we used in this paper are different to those used in \cite{Abbate+2019}, since the publication of new MUSE based cluster velocities from \citet{Martens+2023} has led to a reduction of the HST kinematic distance, this has resulted in a lower value for the best-fitting distance over all methods in \citet{Baumgardt+Vasiliev2021}. 

\subsection{General characteristics of the pulsar population of M62}

\citet{Verbunt+Freire2014} found that the encounter rate for a single binary $\gamma \propto \rho_0^{0.5}r_{\rm c}^{-1}$ is positively and strongly correlated with the fraction of isolated pulsars in a GC. This quantity is also correlated with the presence of slow/high-B-field pulsars (via, hypothetically, the disruption of LMXBs, leaving behind partially recycled pulsars, although other mechanisms might be possible). In addition, $\gamma$ correlates with products of secondary exchange interactions, where an already recycled MSP, or even a partially recycled NS in a LMXB (which in GCs are products of exchange interactions) exchange their former/current donors for new companions. If those are degenerate, then we can observe a MSP in an eccentric orbit with a potentially massive companion; if the new companion is a main sequence or a giant star, then we might instead observe an eclipsing binary, or a partially obscured binary \citep{Zhang_2022}, or, if the obscuration is large, we might not detect a pulsar at all, contributing further to the small fraction of binary pulsars in these clusters.

These empirical trends have been significantly reinforced by recent findings: in the core-collapsed GCs where we know many pulsars reside (at least six pulsars are known in M15, NGC~6752, NGC~6624, NGC~6517, NGC~6522, Terzan~1), the vast majority of previous pulsars and new discoveries are isolated \citep{Ridolfi+2021,Pan+2021,Abbate+2022}, with some of the new discoveries likely being secondary exchange products \citep{Lynch+2012,2015_DeCesar,Ridolfi+2021} and others having high B-fields \citep{Abbate+2022}. However, and somewhat unexpectedly, for the core-collapsed GCs with three or fewer known pulsars (like NGC~6342, 6397, 6544, 6652), most of the pulsars known are in binary systems \citep{Lynch+2012,2015_DeCesar,Gautam_2022,Zhang_2022}.

The dynamical state of M62 has been a subject of debate. This GC has previously been classified as a core-collapsed cluster \citep{Djorgovski+Meylan1993}. However, \citet{Beccari+2006} ruled out this hypothesis based on photometric analysis, where they estimated a core radius $r_{\rm c}= 19$\,arcsec, a concentration $c=1.5$, and a central density of $\log \rho_0 = 5.46$\,\msun\,pc$^{-3}$. They concluded that this high value supported the hypothesis that M62 has experienced (or it is experiencing) a phase of very high production of binary systems while the cluster has not yet undergone core collapse. In the 2010 update of their GC catalogue, \citet{Harris1996} provided $r_{\rm c} = 13$\,arcsec, $c=1.7$, and $\log \rho_{\rm c} = 5.16$\,L$_{\sun}$\,pc$^{-3}$, which are similar to the values obtained by \citet{Beccari+2006}. Using the values from \citet{Harris1996},  \cite{Verbunt+Freire2014} obtained an intermediate value of $\gamma = 9.8$ (scaled to the value of M4). They found that M62 was one of just two clusters that they analysed which seemed to deviate from the correlation of $\gamma$ with the number of isolated pulsars. 

However, the new cluster parameters from Baumgardt's catalogue ($r_{\rm c} = 7.8$\,arcsec, $c=2.35$, and $\log \rho_0 = 5.93$\,\msun\,pc$^{-3}$) represent a smaller core radius and also a higher core density than previously estimated. Using these new values, we obtain $\gamma = 31.14$ (again scaled to the value of M4), which is in the range of the core-collapsed clusters from Table 1 of \citet{Verbunt+Freire2014}.

We can check this further by calculating the ratio of creation and disruption of binaries in this GC, as done by \cite{Beccari+2006}. For the creation of binaries, this was assumed to be the total encounter rate, $\Gamma \propto \rho_0^{1.5}r_{\rm c}^2$ \citep{1987_Verbunt+Hut}, for the disruption, they assumed it to be $\gamma$. For M62, they used the cluster parameters they obtained to derive $\Gamma/\gamma = 2.97$ (normalised to the values of the GC 47 Tuc). Using the new parameters from Baumgardt's catalogue, we obtain $\Gamma/\gamma = 0.5$ (also normalised to the same values of the GC 47 Tuc). These values confirm that the percentage of binaries in M62 should be lower than that observed in 47~Tuc, but also show that this depends crucially on the GC parameters.

If M62 is a core-collapsed GC, its pulsar population is dramatically different from the observed populations of all core-collapsed GCs with many known pulsars.
This finding implies, as discussed in \citet{Possenti+2003}, that the fraction of isolated pulsars might depend, in a way that is not yet understood, on other factors, like the dynamical history of the cluster, especially the detailed evolutionary phase relative to core collapse (for instance, it is possible that in M62 the ongoing process of core collapse has not yet had time to destroy the binaries), the star formation history, the neutron star retention fractions and the orbital period distribution of the binary systems. 
A better understanding of pulsars' populations in GCs might require detailed simulations and further observations. The fact that M62 has many binaries might be related to the fact that in core-collapsed GCs where few pulsars are known, those pulsars are mostly in binary systems.

\subsection{Individual pulsars}

The absence of observed eclipses in the case of M62F could be due to a low orbital inclination for this system. However, the fact that there are also no higher order orbital frequency derivatives as usually seen in such systems (especially with a timing baseline of 23\,yr), suggests that the companion most likely is well inside its Roche Lobe and it is not experiencing tidal deformation (or it is minimal). This might suggest the existence of systems like this, where the companion has shrunk back in its Roche Lobe. If such a configuration is dynamically stable, it could imply that the process of companion ablation in systems like this is not entirely efficient. Consequently, we may anticipate a higher prevalence of such very low mass companion stars in similar systems, and getting isolated MSPs is hard.

The M62H ToAs are well described by a circular orbit and do not show any evidence for eclipses around superior conjunction, similar to the case of M62F. From its timing solution and assuming an inclination angle of 90$^{\circ}$ and a pulsar mass of 1.4\,\msun, we find a minimum companion mass $M_{\rm c,min} = 0.00236$\,\msun ($\sim 2.5$\,M$_{\rm J}$). This value is $\sim 3$ times smaller than the previously known lightest (minimum) mass pulsar companion in a GC, NGC 6440H with $M_{\rm c,min} = 0.006$\,\msun \citep[$\sim 6.6$\,M$_{\rm J}$;][]{Vleeschower+2022}. Also beating its place as the third lowest among all known pulsars\footnote{\url{https://www.atnf.csiro.au/research/pulsar/psrcat/}} according to the ATNF pulsar catalogue \citep[version 1.71, Nov 2023;][]{Manchester+2005}, just after PSR~J2322--2650 \citep[$M_{\rm c,min} = 0.0007$\,\msun, ($\sim 0.7$\,M$_{\rm J}$);][]{2018MNRAS_Spiewak+} and PSR~J1719--1438 \citep[$M_{\rm c,min} = 0.0011$\,\msun ($\sim 1.2$\,M$_{\rm J}$);][]{Bailes+2011}. 

We estimate the Roche Lobe radius of the M62H companion by using Equation 3 in \citet{Bailes+2011} for $i=90^{\circ}$ and a pulsar mass of 1.4\,\msun to be $R_{\rm L} = 3.7 \times 10^{4}$ km, which is about half the radius of Jupiter and 1.3 times the minimum $R_{\rm L}$ for the ultra-low mass carbon white dwarf companion of PSR~J1719--1438. We also calculate a lower limit on the density using the mean density-orbital period relation \citep{Frank+1985}, which is independent of the inclination angle, to be $\rho = 11$\,g\,cm$^{-3}$. This value is about half the density of the ultra-low mass carbon white dwarf companion of PSR~J1719--1438. \citet{vanHaaften+2012} modelled the evolution of ultra-compact X-ray binary systems (UCXBs) concluding that such objects could be the origin of MSPs with very low mass companions, like the case of PSR~J1719--1438. According to Figure 3 in their paper, the most plausible scenario for M62H is that the system could have started as an UCXB of helium composition. \citet{Sengar+2017} uses numerical simulations to study the evolutionary path of UCXBs with a helium white dwarf donor, resulting in a NS with a donor star that becomes a planet with a mass as small as $\sim 0.005$\,\msun~after several Gyr and a radius close to the maximum radius of a cold planet, but with $P_{\rm obs} = 70-80$\,min. This estimation of the mass coincides with the estimated maximum companion mass of M62H, although the orbital period is around 2.4 times shorter. However, according to \citet{vanHaaften+2012}, irradiation effects accelerate the evolution of wider UCXBs with helium white dwarf (He WD) donors and then forming longer orbital periods and lower companion masses. The modelling from \citet{Sengar+2017} accounts for UCXBs with helium of hydrogen lines spectra. Getting optical observations to determine the chemical composition of these MSPs, can help us to disentangle their nature. Recently, rocky exoplanets have been found with similar densities as of M62H, like the case of K2-38b \citep{Toledo-Padron+2020}, with a mean density of around $\rho = 11$\,g\,cm$^{-3}$. The nature of the companion of M62H could then be possible explained as either an ultra-low mass He WD, a primordial planet that could have formed from a disc of material around the pulsar, or as a brown dwarf. This interpretation depends on factors such as the system's inclination and chemical composition.
 
We have only detected M62J in the UHF data despite an in depth and wide ranging search for it in the MeerKAT L-band data. We measure its flux density at UHF $S_{816} = 0.08$\,mJy, and obtained an upper limit of the flux density at L-band assuming the same pulse width and a S/N=3. With this, we obtain a relatively steep spectrum ($\beta < -3.09$). We attribute this and its intrinsic low luminosity from preventing us from detecting this pulsar at L-band frequencies.

The detection of both the eclipsing pulsars during the eclipse phase is an indication that the orbital inclinations are likely not very small and it also implies a highly variable concentration of the eclipsing material. This is particularly evident in the case of the redback pulsar M62B, in which some delays in the ToAs are also seen near inferior conjunction or at 0.81 in phase. These delays have been seen only in one of the observations (UTC 23-07-05-16:57). Material floating around inferior conjunction can be present due to mass transfer during accretion phase through L2 \citep{Linial+2017}. We could not see this at lower frequencies, where the effects of the eclipse are more evident, as our observations did not cover those orbital phases. We suggest future investigations of this pulsar at lower frequencies to study the frequency dependence of the eclipse around inferior conjunction. 

n the case of M62B, we studied the eclipse across four different frequency bands. We observed that the eclipse duration at 612-816 MHz is the longest among the bands, covering 25\% of the orbital phase, followed by the 817--1020\,MHz and 962--1284\,MHz bands with a phase coverage of 22\%, and 19\% for the 1285--1605\,MHz band. We note that both the 817--1020 and 962--1284\,MHz bands have similar eclipse duration since there is an overlap of at least 18\% of the band. Considering that the eclipse is $\sim 1.3$ times longer at the bottom than at the top of the bands, we estimate a power-law index of $-$0.35, similar values have been seen for other black widows \citep{Fruchter+1988, Polzin+2018}. Additionally, we observe that the duration of the egress is more dependent on frequency compared to the ingress.

The total eclipse duration of M62B is $\sim 22$\% of the orbit, while for the case of M62E, it is $\sim 12$\%. Long eclipses (>10\% of the orbital period) at around companion's superior conjunction has been seen in the majority of spider pulsars \citep{Polzin+2020}. Our results provide further evidence supporting the hypothesis that redbacks exhibit relatively longer eclipse duration compared to black widows \citep{Kudale+2020}.


\section{Conclusions}
\label{s:Conclusions}

The high sensitivity of MeerKAT allowed the discovery of three new pulsars in M62, including the very faint and relatively steep spectrum ($\beta < - 3.09$) M62J, which was only detected at UHF frequencies. These discoveries bring the total number of pulsars in M62 to ten, with none found to be isolated up to date.  M62H is in a binary with the lightest companion known to date in a GC, with a minimum mass of $2.5$\,$M_{\rm J}$. The system does not display eclipses in the MeerKAT data.

The fact that only binary pulsars have been found in M62 is surprising, being fundamentally at odds with the observed characteristics of the pulsar populations of other GCs. This opens the door for further investigation of the relation between the dynamical properties and history of GCs and the characteristics of their pulsar populations.

The large bandwidth of MeerKAT enabled us to carry out an analysis of the eclipses of the two eclipsing systems in the cluster, M62B and M62E, a redback and a black widow pulsar, respectively. These analyses reveal a clear evidence of frequency dependence, with longer and asymmetric eclipses occurring at lower frequencies. 

The large timing baseline provided by both Parkes and MeerKAT has produced a significant improvement in the measurement of several parameters for the previously known pulsars, including the description of the behaviour for M62B and M62E with a large number of orbital frequency derivatives. We present several new additional pulsar parameters that are important for a study of the dynamics of the cluster: the proper motions, the real line-of-sight accelerations as determined from the orbital period derivatives and the jerks. These should be helpful for more detailed analyses such as that previously done by \citet{Prager+2017} and will be able to support or contradict the evidence of non-luminous mass at the centre of the cluster \citep{Abbate+2019}.

\section*{Acknowledgements}
The MeerKAT telescope is operated by the South African Radio Astronomy Observatory, which is a facility of the National Research Foundation, an agency of the Department of Science and Innovation. SARAO acknowledges the ongoing advice and calibration of GPS systems by the National Metrology Institute of South Africa (NMISA) and the time space reference systems department of the Paris Observatory. MeerTime data is housed on the OzSTAR supercomputer at Swinburne University of Technology. The OzSTAR program receives funding in part from the Astronomy National Collaborative Research Infrastructure Strategy (NCRIS) allocation provided by the Australian Government.
The Parkes radio telescope is funded by the Commonwealth of Australia for operation as a National Facility
managed by CSIRO. We acknowledge the Wiradjuri people as the traditional owners of the Observatory site. 
The Green Bank Observatory is a facility of the National Science Foundation operated under cooperative agreement by Associated Universities, Inc.
The National Radio Astronomy Observatory (NRAO) is a facility of the National Science Foundation operated under cooperative agreement by Associated Universities, Inc. 
The authors also acknowledge MPIfR funding to contribute to MeerTime infrastructure. TRAPUM observations used the FBFUSE and APSUSE computing clusters for data acquisition, storage and analysis. These clusters were funded and installed by the Max-Planck-Institut für Radioastronomie and the Max-Planck-Gesellschaft. 
LV acknowledges financial support from the Dean’s Doctoral Scholar Award from the University of Manchester. 
BWS acknowledges funding from the European Research Council (ERC) under the European Union’s Horizon 2020 research and innovation programme (grant agreement No. 694745).
Part of this work has been funded using resources from the INAF Large Grant 2022 “GCjewels” (P.I. Andrea Possenti) approved with the Presidential Decree 30/2022. This work was also in part supported by the “Italian Ministry of Foreign Affairs and International Cooperation”, grant number ZA23GR03.
PCCF, AR, PVP, VB, MK, VVK, EDB, and WC acknowledge continuing valuable support from the Max-Planck Society. 
SMR is a CIFAR Fellow and is supported by the NSF Physics Frontiers Center award 2020265.
VVK acknowledges financial support from the European Research Council (ERC) starting grant ,"COMPACT" (Grant agreement number 101078094), under the European Union's Horizon Europe research and innovation programme. 
MB acknowledges support through ARC grant CE170100004.
LZ acknowledges financial support from ACAMAR Postdoctoral Fellowship and the National Natural Science Foundation of China (Grant No. 12103069). 
\\
The authors acknowledge H. Baumgardt for useful discussions about the parameters of M62.
We thank the anonymous referee for their helpful comments, and constructive remarks on this manuscript.

\section*{Data Availability}
The data that support the findings of this article will be shared on reasonable request to the MeerTime and TRAPUM collaborations.




\bibliographystyle{mnras}
\bibliography{NGC6266} 

\begin{thebibliography}{}
\makeatletter
\relax
\def\mn@urlcharsother{\let\do\@makeother \do\$\do\&\do\#\do\^\do\_\do\%\do\~}
\def\mn@doi{\begingroup\mn@urlcharsother \@ifnextchar [ {\mn@doi@}
  {\mn@doi@[]}}
\def\mn@doi@[#1]#2{\def\@tempa{#1}\ifx\@tempa\@empty \href
  {http://dx.doi.org/#2} {doi:#2}\else \href {http://dx.doi.org/#2} {#1}\fi
  \endgroup}
\def\mn@eprint#1#2{\mn@eprint@#1:#2::\@nil}
\def\mn@eprint@arXiv#1{\href {http://arxiv.org/abs/#1} {{\tt arXiv:#1}}}
\def\mn@eprint@dblp#1{\href {http://dblp.uni-trier.de/rec/bibtex/#1.xml}
  {dblp:#1}}
\def\mn@eprint@#1:#2:#3:#4\@nil{\def\@tempa {#1}\def\@tempb {#2}\def\@tempc
  {#3}\ifx \@tempc \@empty \let \@tempc \@tempb \let \@tempb \@tempa \fi \ifx
  \@tempb \@empty \def\@tempb {arXiv}\fi \@ifundefined
  {mn@eprint@\@tempb}{\@tempb:\@tempc}{\expandafter \expandafter \csname
  mn@eprint@\@tempb\endcsname \expandafter{\@tempc}}}

\bibitem[\protect\citeauthoryear{{Abbate}, {Possenti}, {Ridolfi}, {Freire},
  {Camilo}, {Manchester}  \& {D'Amico}}{{Abbate} et~al.}{2018}]{Abbate+2018}
{Abbate} F.,  {Possenti} A.,  {Ridolfi} A.,  {Freire} P.~C.~C.,  {Camilo} F.,
  {Manchester} R.~N.,   {D'Amico} N.,  2018, \mn@doi [\mnras]
  {10.1093/mnras/sty2298}, \href
  {https://ui.adsabs.harvard.edu/abs/2018MNRAS.481..627A} {481, 627}

\bibitem[\protect\citeauthoryear{{Abbate}, {Possenti}, {Colpi}  \&
  {Spera}}{{Abbate} et~al.}{2019}]{Abbate+2019}
{Abbate} F.,  {Possenti} A.,  {Colpi} M.,   {Spera} M.,  2019, \mn@doi [\apjl]
  {10.3847/2041-8213/ab46c3}, \href
  {https://ui.adsabs.harvard.edu/abs/2019ApJ...884L...9A} {884, L9}

\bibitem[\protect\citeauthoryear{{Abbate} et~al.,}{{Abbate}
  et~al.}{2022}]{Abbate+2022}
{Abbate} F.,  et~al., 2022, \mn@doi [\mnras] {10.1093/mnras/stac1041}, \href
  {https://ui.adsabs.harvard.edu/abs/2022MNRAS.513.2292A} {513, 2292}

\bibitem[\protect\citeauthoryear{{Andersen} \& {Ransom}}{{Andersen} \&
  {Ransom}}{2018}]{Andersen+2018}
{Andersen} B.~C.,  {Ransom} S.~M.,  2018, \mn@doi [\apjl]
  {10.3847/2041-8213/aad59f}, \href
  {https://ui.adsabs.harvard.edu/abs/2018ApJ...863L..13A} {863, L13}

\bibitem[\protect\citeauthoryear{{Bailes} et~al.,}{{Bailes}
  et~al.}{2011}]{Bailes+2011}
{Bailes} M.,  et~al., 2011, \mn@doi [Science] {10.1126/science.1208890}, \href
  {https://ui.adsabs.harvard.edu/abs/2011Sci...333.1717B} {333, 1717}

\bibitem[\protect\citeauthoryear{{Bailes} et~al.,}{{Bailes}
  et~al.}{2020}]{Bailes+2020}
{Bailes} M.,  et~al., 2020, \mn@doi [\pasa] {10.1017/pasa.2020.19}, \href
  {https://ui.adsabs.harvard.edu/abs/2020PASA...37...28B} {37, e028}

\bibitem[\protect\citeauthoryear{{Barr}}{{Barr}}{2018}]{Barr2018}
{Barr} E.~D.,  2018, in {Weltevrede} P.,  {Perera} B.~B.~P.,  {Preston} L.~L.,
   {Sanidas} S.,  eds,  IAU symposium and colloquium proceedings series Vol.
  337, Pulsar Astrophysics the Next Fifty Years. pp 175--178,
  \mn@doi{10.1017/S1743921317009036}

\bibitem[\protect\citeauthoryear{{Baumgardt} \& {Hilker}}{{Baumgardt} \&
  {Hilker}}{2018}]{Baumgardt+Hilker2018}
{Baumgardt} H.,  {Hilker} M.,  2018, \mn@doi [\mnras] {10.1093/mnras/sty1057},
  \href {https://ui.adsabs.harvard.edu/abs/2018MNRAS.478.1520B} {478, 1520}

\bibitem[\protect\citeauthoryear{{Baumgardt} \& {Vasiliev}}{{Baumgardt} \&
  {Vasiliev}}{2021}]{Baumgardt+Vasiliev2021}
{Baumgardt} H.,  {Vasiliev} E.,  2021, \mn@doi [\mnras]
  {10.1093/mnras/stab1474}, \href
  {https://ui.adsabs.harvard.edu/abs/2021MNRAS.505.5957B} {505, 5957}

\bibitem[\protect\citeauthoryear{{Beccari}, {Ferraro}, {Possenti}, {Valenti},
  {Origlia}  \& {Rood}}{{Beccari} et~al.}{2006}]{Beccari+2006}
{Beccari} G.,  {Ferraro} F.~R.,  {Possenti} A.,  {Valenti} E.,  {Origlia} L.,
  {Rood} R.~T.,  2006, \mn@doi [\aj] {10.1086/500643}, \href
  {https://ui.adsabs.harvard.edu/abs/2006AJ....131.2551B} {131, 2551}

\bibitem[\protect\citeauthoryear{{Bezuidenhout}, {Clark}, {Breton}  \&
  {Stappers}}{{Bezuidenhout} et~al.}{2023}]{Bezuidenhout+2023}
{Bezuidenhout} M.~C.,  {Clark} C.~J.,  {Breton} R.~P.,   {Stappers} B.~W.,
  2023, {SeeKAT: Localizer for transients detected in tied-array beams},
  Astrophysics Source Code Library, record ascl:2303.003 (\mn@eprint {ascl}
  {2303.003})

\bibitem[\protect\citeauthoryear{{Blandford}, {Romani}  \&
  {Applegate}}{{Blandford} et~al.}{1987}]{Blandford+1987}
{Blandford} R.~D.,  {Romani} R.~W.,   {Applegate} J.~H.,  1987, \mn@doi
  [\mnras] {10.1093/mnras/225.1.51P}, \href
  {https://ui.adsabs.harvard.edu/abs/1987MNRAS.225P..51B} {225, 51P}

\bibitem[\protect\citeauthoryear{{Booth} \& {Jonas}}{{Booth} \&
  {Jonas}}{2012}]{Booth+Jonas2012}
{Booth} R.~S.,  {Jonas} J.~L.,  2012, African Skies, \href
  {https://ui.adsabs.harvard.edu/abs/2012AfrSk..16..101B} {16, 101}

\bibitem[\protect\citeauthoryear{{Chen}, {Barr}, {Karuppusamy}, {Kramer}  \&
  {Stappers}}{{Chen} et~al.}{2021}]{Chen+2021}
{Chen} W.,  {Barr} E.,  {Karuppusamy} R.,  {Kramer} M.,   {Stappers} B.,  2021,
  arXiv e-prints, \href {https://ui.adsabs.harvard.edu/abs/2021arXiv211001667C}
  {p. arXiv:2110.01667}

\bibitem[\protect\citeauthoryear{{Chomiuk}, {Strader}, {Maccarone},
  {Miller-Jones}, {Heinke}, {Noyola}, {Seth}  \& {Ransom}}{{Chomiuk}
  et~al.}{2013}]{Chomiuk+2013}
{Chomiuk} L.,  {Strader} J.,  {Maccarone} T.~J.,  {Miller-Jones} J. C.~A.,
  {Heinke} C.,  {Noyola} E.,  {Seth} A.~C.,   {Ransom} S.,  2013, \mn@doi
  [\apj] {10.1088/0004-637X/777/1/69}, \href
  {https://ui.adsabs.harvard.edu/abs/2013ApJ...777...69C} {777, 69}

\bibitem[\protect\citeauthoryear{{Cocozza}, {Ferraro}, {Possenti}, {Beccari},
  {Lanzoni}, {Ransom}, {Rood}  \& {D'Amico}}{{Cocozza}
  et~al.}{2008}]{Cocozza+2008}
{Cocozza} G.,  {Ferraro} F.~R.,  {Possenti} A.,  {Beccari} G.,  {Lanzoni} B.,
  {Ransom} S.,  {Rood} R.~T.,   {D'Amico} N.,  2008, \mn@doi [\apjl]
  {10.1086/589557}, \href
  {https://ui.adsabs.harvard.edu/abs/2008ApJ...679L.105C} {679, L105}

\bibitem[\protect\citeauthoryear{{Cordes} \& {Lazio}}{{Cordes} \&
  {Lazio}}{2002}]{NE2001}
{Cordes} J.~M.,  {Lazio} T.~J.~W.,  2002, \mn@doi [arXiv e-prints]
  {10.48550/arXiv.astro-ph/0207156}, \href
  {https://ui.adsabs.harvard.edu/abs/2002astro.ph..7156C} {pp
  astro--ph/0207156}

\bibitem[\protect\citeauthoryear{{D'Amico}, {Lyne}, {Manchester}, {Possenti}
  \& {Camilo}}{{D'Amico} et~al.}{2001a}]{DAmico+2001}
{D'Amico} N.,  {Lyne} A.~G.,  {Manchester} R.~N.,  {Possenti} A.,   {Camilo}
  F.,  2001a, \mn@doi [\apjl] {10.1086/319096}, \href
  {https://ui.adsabs.harvard.edu/abs/2001ApJ...548L.171D} {548, L171}

\bibitem[\protect\citeauthoryear{{D'Amico}, {Possenti}, {Manchester},
  {Sarkissian}, {Lyne}  \& {Camilo}}{{D'Amico} et~al.}{2001b}]{DAmico+2001b}
{D'Amico} N.,  {Possenti} A.,  {Manchester} R.~N.,  {Sarkissian} J.,  {Lyne}
  A.~G.,   {Camilo} F.,  2001b, in {Wheeler} J.~C.,  {Martel} H.,  eds,
  American Institute of Physics Conference Series Vol. 586, 20th Texas
  Symposium on relativistic astrophysics. pp 526--531 (\mn@eprint {arXiv}
  {astro-ph/0105122}), \mn@doi{10.1063/1.1419604}

\bibitem[\protect\citeauthoryear{{Damour} \& {Taylor}}{{Damour} \&
  {Taylor}}{1991}]{Damour+Taylor1991}
{Damour} T.,  {Taylor} J.~H.,  1991, \mn@doi [\apj] {10.1086/169585}, \href
  {https://ui.adsabs.harvard.edu/abs/1991ApJ...366..501D} {366, 501}

\bibitem[\protect\citeauthoryear{{DeCesar}, {Ransom}, {Kaplan}, {Ray}  \&
  {Geller}}{{DeCesar} et~al.}{2015}]{2015_DeCesar}
{DeCesar} M.~E.,  {Ransom} S.~M.,  {Kaplan} D.~L.,  {Ray} P.~S.,   {Geller}
  A.~M.,  2015, \mn@doi [\apjl] {10.1088/2041-8205/807/2/L23}, \href
  {https://ui.adsabs.harvard.edu/abs/2015ApJ...807L..23D} {807, L23}

\bibitem[\protect\citeauthoryear{{Dewey}, {Taylor}, {Weisberg}  \&
  {Stokes}}{{Dewey} et~al.}{1985}]{Dewey+1985}
{Dewey} R.~J.,  {Taylor} J.~H.,  {Weisberg} J.~M.,   {Stokes} G.~H.,  1985,
  \mn@doi [\apjl] {10.1086/184502}, \href
  {https://ui.adsabs.harvard.edu/abs/1985ApJ...294L..25D} {294, L25}

\bibitem[\protect\citeauthoryear{{Djorgovski} \& {Meylan}}{{Djorgovski} \&
  {Meylan}}{1993}]{Djorgovski+Meylan1993}
{Djorgovski} S.,  {Meylan} G.,  1993, in American Astronomical Society Meeting
  Abstracts \#182. p. 50.15

\bibitem[\protect\citeauthoryear{{Folkner}, {Williams}  \& {Boggs}}{{Folkner}
  et~al.}{2009}]{Folkner+2009}
{Folkner} W.~M.,  {Williams} J.~G.,   {Boggs} D.~H.,  2009, Interplanetary
  Network Progress Report, \href
  {https://ui.adsabs.harvard.edu/abs/2009IPNPR.178C...1F} {42-178, 1}

\bibitem[\protect\citeauthoryear{{Frank}, {King}  \& {Raine}}{{Frank}
  et~al.}{1985}]{Frank+1985}
{Frank} J.,  {King} A.~R.,   {Raine} D.~J.,  1985, {Accretion power in
  astrophysics}.
Cambridge University Press

\bibitem[\protect\citeauthoryear{{Freire} \& {Ridolfi}}{{Freire} \&
  {Ridolfi}}{2018}]{Freire+Ridolfi2018}
{Freire} P. C.~C.,  {Ridolfi} A.,  2018, \mn@doi [\mnras]
  {10.1093/mnras/sty524}, \href
  {https://ui.adsabs.harvard.edu/abs/2018MNRAS.476.4794F} {476, 4794}

\bibitem[\protect\citeauthoryear{{Freire}, {Kramer}  \& {Lyne}}{{Freire}
  et~al.}{2001a}]{2001MNRAS_F}
{Freire} P.~C.,  {Kramer} M.,   {Lyne} A.~G.,  2001a, \mn@doi [\mnras]
  {10.1046/j.1365-8711.2001.04200.x}, \href
  {https://ui.adsabs.harvard.edu/abs/2001MNRAS.322..885F} {322, 885}

\bibitem[\protect\citeauthoryear{{Freire}, {Kramer}, {Lyne}, {Camilo},
  {Manchester}  \& {D'Amico}}{{Freire} et~al.}{2001b}]{Freire+2001}
{Freire} P.~C.,  {Kramer} M.,  {Lyne} A.~G.,  {Camilo} F.,  {Manchester} R.~N.,
    {D'Amico} N.,  2001b, \mn@doi [\apjl] {10.1086/323248}, \href
  {https://ui.adsabs.harvard.edu/abs/2001ApJ...557L.105F} {557, L105}

\bibitem[\protect\citeauthoryear{{Freire}, {Ransom}, {B{\'e}gin}, {Stairs},
  {Hessels}, {Frey}  \& {Camilo}}{{Freire} et~al.}{2008a}]{Freire+2008a}
{Freire} P. C.~C.,  {Ransom} S.~M.,  {B{\'e}gin} S.,  {Stairs} I.~H.,
  {Hessels} J. W.~T.,  {Frey} L.~H.,   {Camilo} F.,  2008a, \mn@doi [\apj]
  {10.1086/526338}, \href
  {https://ui.adsabs.harvard.edu/abs/2008ApJ...675..670F} {675, 670}

\bibitem[\protect\citeauthoryear{{Freire}, {Wolszczan}, {van den Berg}  \&
  {Hessels}}{{Freire} et~al.}{2008b}]{Freire+2008b}
{Freire} P. C.~C.,  {Wolszczan} A.,  {van den Berg} M.,   {Hessels} J. W.~T.,
  2008b, \mn@doi [\apj] {10.1086/587832}, \href
  {https://ui.adsabs.harvard.edu/abs/2008ApJ...679.1433F} {679, 1433}

\bibitem[\protect\citeauthoryear{{Freire} et~al.,}{{Freire}
  et~al.}{2017}]{Freire+2017}
{Freire} P.~C.~C.,  et~al., 2017, \mn@doi [\mnras] {10.1093/mnras/stx1533},
  \href {https://ui.adsabs.harvard.edu/abs/2017MNRAS.471..857F} {471, 857}

\bibitem[\protect\citeauthoryear{{Fruchter}, {Gunn}, {Lauer}  \&
  {Dressler}}{{Fruchter} et~al.}{1988}]{Fruchter+1988}
{Fruchter} A.~S.,  {Gunn} J.~E.,  {Lauer} T.~R.,   {Dressler} A.,  1988,
  \mn@doi [\nat] {10.1038/334686a0}, \href
  {https://ui.adsabs.harvard.edu/abs/1988Natur.334..686F} {334, 686}

\bibitem[\protect\citeauthoryear{{GRAVITY Collaboration} et~al.,}{{GRAVITY
  Collaboration} et~al.}{2021}]{Grav+2021}
{GRAVITY Collaboration} et~al., 2021, \mn@doi [\aap]
  {10.1051/0004-6361/202040208}, \href
  {https://ui.adsabs.harvard.edu/abs/2021A&A...647A..59G} {647, A59}

\bibitem[\protect\citeauthoryear{{Gautam} et~al.,}{{Gautam}
  et~al.}{2022}]{Gautam_2022}
{Gautam} T.,  et~al., 2022, \mn@doi [\aap] {10.1051/0004-6361/202243062}, \href
  {https://ui.adsabs.harvard.edu/abs/2022A&A...664A..54G} {664, A54}

\bibitem[\protect\citeauthoryear{{Guo} et~al.,}{{Guo} et~al.}{2021}]{Guo+2021}
{Guo} Y.~J.,  et~al., 2021, \mn@doi [\aap] {10.1051/0004-6361/202141450}, \href
  {https://ui.adsabs.harvard.edu/abs/2021A&A...654A..16G} {654, A16}

\bibitem[\protect\citeauthoryear{{Harris}}{{Harris}}{1996}]{Harris1996}
{Harris} W.~E.,  1996, \mn@doi [\aj] {10.1086/118116}, \href
  {https://ui.adsabs.harvard.edu/abs/1996AJ....112.1487H} {112, 1487}

\bibitem[\protect\citeauthoryear{{Hobbs}, {Edwards}  \& {Manchester}}{{Hobbs}
  et~al.}{2006}]{Hobbs+2006}
{Hobbs} G.,  {Edwards} R.,   {Manchester} R.,  2006, Chinese Journal of
  Astronomy and Astrophysics Supplement, \href
  {https://ui.adsabs.harvard.edu/abs/2006ChJAS...6b.189H} {6, 189}

\bibitem[\protect\citeauthoryear{{Hotan}, {van Straten}  \&
  {Manchester}}{{Hotan} et~al.}{2004}]{Hotan+2004}
{Hotan} A.~W.,  {van Straten} W.,   {Manchester} R.~N.,  2004, \mn@doi [\pasa]
  {10.1071/AS04022}, \href
  {https://ui.adsabs.harvard.edu/abs/2004PASA...21..302H} {21, 302}

\bibitem[\protect\citeauthoryear{{Jacoby}, {Chandler}, {Backer}, {Anderson}  \&
  {Kulkarni}}{{Jacoby} et~al.}{2002}]{Jacoby+2002}
{Jacoby} B.~A.,  {Chandler} A.~M.,  {Backer} D.~C.,  {Anderson} S.~B.,
  {Kulkarni} S.~R.,  2002, \iaucirc, \href
  {https://ui.adsabs.harvard.edu/abs/2002IAUC.7783....1J} {7783, 1}

\bibitem[\protect\citeauthoryear{{Jacoby}, {Cameron}, {Jenet}, {Anderson},
  {Murty}  \& {Kulkarni}}{{Jacoby} et~al.}{2006}]{Jacoby+2006}
{Jacoby} B.~A.,  {Cameron} P.~B.,  {Jenet} F.~A.,  {Anderson} S.~B.,  {Murty}
  R.~N.,   {Kulkarni} S.~R.,  2006, \mn@doi [\apjl] {10.1086/505742}, \href
  {https://ui.adsabs.harvard.edu/abs/2006ApJ...644L.113J} {644, L113}

\bibitem[\protect\citeauthoryear{{Joshi} \& {Rasio}}{{Joshi} \&
  {Rasio}}{1997}]{Joshi+Rasio1997}
{Joshi} K.~J.,  {Rasio} F.~A.,  1997, \mn@doi [\apj] {10.1086/303916}, \href
  {https://ui.adsabs.harvard.edu/abs/1997ApJ...479..948J} {479, 948}

\bibitem[\protect\citeauthoryear{{Kudale}, {Roy}, {Bhattacharyya}, {Stappers}
  \& {Chengalur}}{{Kudale} et~al.}{2020}]{Kudale+2020}
{Kudale} S.,  {Roy} J.,  {Bhattacharyya} B.,  {Stappers} B.,   {Chengalur} J.,
  2020, \mn@doi [\apj] {10.3847/1538-4357/aba902}, \href
  {https://ui.adsabs.harvard.edu/abs/2020ApJ...900..194K} {900, 194}

\bibitem[\protect\citeauthoryear{{Lange}, {Camilo}, {Wex}, {Kramer}, {Backer},
  {Lyne}  \& {Doroshenko}}{{Lange} et~al.}{2001}]{Lange+2001}
{Lange} C.,  {Camilo} F.,  {Wex} N.,  {Kramer} M.,  {Backer} D.~C.,  {Lyne}
  A.~G.,   {Doroshenko} O.,  2001, \mn@doi [\mnras]
  {10.1046/j.1365-8711.2001.04606.x}, \href
  {https://ui.adsabs.harvard.edu/abs/2001MNRAS.326..274L} {326, 274}

\bibitem[\protect\citeauthoryear{{Lapenna}, {Mucciarelli}, {Ferraro},
  {Origlia}, {Lanzoni}, {Massari}  \& {Dalessandro}}{{Lapenna}
  et~al.}{2015}]{Lapenna+2015}
{Lapenna} E.,  {Mucciarelli} A.,  {Ferraro} F.~R.,  {Origlia} L.,  {Lanzoni}
  B.,  {Massari} D.,   {Dalessandro} E.,  2015, \mn@doi [\apj]
  {10.1088/0004-637X/813/2/97}, \href
  {https://ui.adsabs.harvard.edu/abs/2015ApJ...813...97L} {813, 97}

\bibitem[\protect\citeauthoryear{{Lazaridis} et~al.,}{{Lazaridis}
  et~al.}{2009}]{Lazaridis+2009}
{Lazaridis} K.,  et~al., 2009, \mn@doi [\mnras]
  {10.1111/j.1365-2966.2009.15481.x}, \href
  {https://ui.adsabs.harvard.edu/abs/2009MNRAS.400..805L} {400, 805}

\bibitem[\protect\citeauthoryear{{Linial} \& {Sari}}{{Linial} \&
  {Sari}}{2017}]{Linial+2017}
{Linial} I.,  {Sari} R.,  2017, \mn@doi [\mnras] {10.1093/mnras/stx1041}, \href
  {https://ui.adsabs.harvard.edu/abs/2017MNRAS.469.2441L} {469, 2441}

\bibitem[\protect\citeauthoryear{{Lynch}, {Freire}, {Ransom}  \&
  {Jacoby}}{{Lynch} et~al.}{2012}]{Lynch+2012}
{Lynch} R.~S.,  {Freire} P. C.~C.,  {Ransom} S.~M.,   {Jacoby} B.~A.,  2012,
  \mn@doi [\apj] {10.1088/0004-637X/745/2/109}, \href
  {https://ui.adsabs.harvard.edu/abs/2012ApJ...745..109L} {745, 109}

\bibitem[\protect\citeauthoryear{{Manchester}, {Hobbs}, {Teoh}  \&
  {Hobbs}}{{Manchester} et~al.}{2005}]{Manchester+2005}
{Manchester} R.~N.,  {Hobbs} G.~B.,  {Teoh} A.,   {Hobbs} M.,  2005, VizieR
  Online Data Catalog, \href
  {https://ui.adsabs.harvard.edu/abs/2005yCat.7245....0M} {p. VII/245}

\bibitem[\protect\citeauthoryear{{Martens} et~al.,}{{Martens}
  et~al.}{2023}]{Martens+2023}
{Martens} S.,  et~al., 2023, \mn@doi [\aap] {10.1051/0004-6361/202244787},
  \href {https://ui.adsabs.harvard.edu/abs/2023A&A...671A.106M} {671, A106}

\bibitem[\protect\citeauthoryear{{Men}, {Barr}, {Clark}, {Carli}  \&
  {Desvignes}}{{Men} et~al.}{2023}]{Men+2023}
{Men} Y.,  {Barr} E.,  {Clark} C.~J.,  {Carli} E.,   {Desvignes} G.,  2023,
  \mn@doi [arXiv e-prints] {10.48550/arXiv.2309.02544}, \href
  {https://ui.adsabs.harvard.edu/abs/2023arXiv230902544M} {p. arXiv:2309.02544}

\bibitem[\protect\citeauthoryear{{Meylan} \& {Heggie}}{{Meylan} \&
  {Heggie}}{1997}]{Meylan+Heggie1997}
{Meylan} G.,  {Heggie} D.~C.,  1997, \mn@doi [\aapr] {10.1007/s001590050008},
  \href {https://ui.adsabs.harvard.edu/abs/1997A&ARv...8....1M} {8, 1}

\bibitem[\protect\citeauthoryear{{Morello}, {Rajwade}  \& {Stappers}}{{Morello}
  et~al.}{2022}]{Morello+2022}
{Morello} V.,  {Rajwade} K.~M.,   {Stappers} B.~W.,  2022, \mn@doi [\mnras]
  {10.1093/mnras/stab3493}, \href
  {https://ui.adsabs.harvard.edu/abs/2022MNRAS.510.1393M} {510, 1393}

\bibitem[\protect\citeauthoryear{{Nice} et~al.,}{{Nice}
  et~al.}{2015}]{Nice+2015}
{Nice} D.,  et~al., 2015, {Tempo: Pulsar timing data analysis}, Astrophysics
  Source Code Library, record ascl:1509.002

\bibitem[\protect\citeauthoryear{{Oh}, {Hui}, {Li}  \& {Kong}}{{Oh}
  et~al.}{2020}]{Oh+2020}
{Oh} K.,  {Hui} C.~Y.,  {Li} K.~L.,   {Kong} A.~K.~H.,  2020, \mn@doi [\mnras]
  {10.1093/mnras/staa2462}, \href
  {https://ui.adsabs.harvard.edu/abs/2020MNRAS.498..292O} {498, 292}

\bibitem[\protect\citeauthoryear{{Pan} et~al.,}{{Pan} et~al.}{2021}]{Pan+2021}
{Pan} Z.,  et~al., 2021, \mn@doi [\apjl] {10.3847/2041-8213/ac0bbd}, \href
  {https://ui.adsabs.harvard.edu/abs/2021ApJ...915L..28P} {915, L28}

\bibitem[\protect\citeauthoryear{{Phinney}}{{Phinney}}{1992}]{Phinney1992}
{Phinney} E.~S.,  1992, \mn@doi [Philosophical Transactions of the Royal
  Society of London Series A] {10.1098/rsta.1992.0084}, \href
  {https://ui.adsabs.harvard.edu/abs/1992RSPTA.341...39P} {341, 39}

\bibitem[\protect\citeauthoryear{{Phinney}}{{Phinney}}{1993}]{Phinney1993}
{Phinney} E.~S.,  1993, in {Djorgovski} S.~G.,  {Meylan} G.,  eds,
  Astronomical Society of the Pacific Conference Series Vol. 50, Structure and
  Dynamics of Globular Clusters. p.~141

\bibitem[\protect\citeauthoryear{{Polzin} et~al.,}{{Polzin}
  et~al.}{2018}]{Polzin+2018}
{Polzin} E.~J.,  et~al., 2018, \mn@doi [\mnras] {10.1093/mnras/sty349}, \href
  {https://ui.adsabs.harvard.edu/abs/2018MNRAS.476.1968P} {476, 1968}

\bibitem[\protect\citeauthoryear{{Polzin}, {Breton}, {Bhattacharyya},
  {Scholte}, {Sobey}  \& {Stappers}}{{Polzin} et~al.}{2020}]{Polzin+2020}
{Polzin} E.~J.,  {Breton} R.~P.,  {Bhattacharyya} B.,  {Scholte} D.,  {Sobey}
  C.,   {Stappers} B.~W.,  2020, \mn@doi [\mnras] {10.1093/mnras/staa596},
  \href {https://ui.adsabs.harvard.edu/abs/2020MNRAS.494.2948P} {494, 2948}

\bibitem[\protect\citeauthoryear{{Pooley} et~al.,}{{Pooley}
  et~al.}{2003}]{Pooley+2003}
{Pooley} D.,  et~al., 2003, \mn@doi [\apjl] {10.1086/377074}, \href
  {https://ui.adsabs.harvard.edu/abs/2003ApJ...591L.131P} {591, L131}

\bibitem[\protect\citeauthoryear{{Possenti}, {D'Amico}, {Manchester}, {Camilo},
  {Lyne}, {Sarkissian}  \& {Corongiu}}{{Possenti} et~al.}{2003}]{Possenti+2003}
{Possenti} A.,  {D'Amico} N.,  {Manchester} R.~N.,  {Camilo} F.,  {Lyne} A.~G.,
   {Sarkissian} J.,   {Corongiu} A.,  2003, \mn@doi [\apj] {10.1086/379190},
  \href {https://ui.adsabs.harvard.edu/abs/2003ApJ...599..475P} {599, 475}

\bibitem[\protect\citeauthoryear{{Prager}, {Ransom}, {Freire}, {Hessels},
  {Stairs}, {Arras}  \& {Cadelano}}{{Prager} et~al.}{2017}]{Prager+2017}
{Prager} B.~J.,  {Ransom} S.~M.,  {Freire} P. C.~C.,  {Hessels} J. W.~T.,
  {Stairs} I.~H.,  {Arras} P.,   {Cadelano} M.,  2017, \mn@doi [\apj]
  {10.3847/1538-4357/aa7ed7}, \href
  {https://ui.adsabs.harvard.edu/abs/2017ApJ...845..148P} {845, 148}

\bibitem[\protect\citeauthoryear{{Ransom}}{{Ransom}}{2001}]{Ransom2001}
{Ransom} S.~M.,  2001, PhD thesis, Harvard University, Massachusetts

\bibitem[\protect\citeauthoryear{{Ransom}, {Eikenberry}  \&
  {Middleditch}}{{Ransom} et~al.}{2002}]{Ransom+2002}
{Ransom} S.~M.,  {Eikenberry} S.~S.,   {Middleditch} J.,  2002, \mn@doi [\aj]
  {10.1086/342285}, \href {http://adsabs.harvard.edu/abs/2002AJ....124.1788R}
  {124, 1788}

\bibitem[\protect\citeauthoryear{{Ridolfi} et~al.,}{{Ridolfi}
  et~al.}{2016}]{Ridolfi+2016}
{Ridolfi} A.,  et~al., 2016, \mn@doi [\mnras] {10.1093/mnras/stw1850}, \href
  {https://ui.adsabs.harvard.edu/abs/2016MNRAS.462.2918R} {462, 2918}

\bibitem[\protect\citeauthoryear{{Ridolfi} et~al.,}{{Ridolfi}
  et~al.}{2021}]{Ridolfi+2021}
{Ridolfi} A.,  et~al., 2021, \mn@doi [\mnras] {10.1093/mnras/stab790}, \href
  {https://ui.adsabs.harvard.edu/abs/2021MNRAS.tmp..783R} {}

\bibitem[\protect\citeauthoryear{{Sengar}, {Tauris}, {Langer}  \&
  {Istrate}}{{Sengar} et~al.}{2017}]{Sengar+2017}
{Sengar} R.,  {Tauris} T.~M.,  {Langer} N.,   {Istrate} A.~G.,  2017, \mn@doi
  [\mnras] {10.1093/mnrasl/slx064}, \href
  {https://ui.adsabs.harvard.edu/abs/2017MNRAS.470L...6S} {470, L6}

\bibitem[\protect\citeauthoryear{{Shaifullah} et~al.,}{{Shaifullah}
  et~al.}{2016}]{Shaifullah+2016}
{Shaifullah} G.,  et~al., 2016, \mn@doi [\mnras] {10.1093/mnras/stw1737}, \href
  {https://ui.adsabs.harvard.edu/abs/2016MNRAS.462.1029S} {462, 1029}

\bibitem[\protect\citeauthoryear{{Shklovskii}}{{Shklovskii}}{1970}]{Shklovskii1970}
{Shklovskii} I.~S.,  1970, \sovast, \href
  {https://ui.adsabs.harvard.edu/abs/1970SvA....13..562S} {13, 562}

\bibitem[\protect\citeauthoryear{{Spiewak} et~al.,}{{Spiewak}
  et~al.}{2018}]{2018MNRAS_Spiewak+}
{Spiewak} R.,  et~al., 2018, \mn@doi [\mnras] {10.1093/mnras/stx3157}, \href
  {https://ui.adsabs.harvard.edu/abs/2018MNRAS.475..469S} {475, 469}

\bibitem[\protect\citeauthoryear{{Stappers} \& {Kramer}}{{Stappers} \&
  {Kramer}}{2016}]{Stappers+Kramer2016}
{Stappers} B.,  {Kramer} M.,  2016, in MeerKAT Science: On the Pathway to the
  SKA. p.~9

\bibitem[\protect\citeauthoryear{{Stappers} et~al.,}{{Stappers}
  et~al.}{1996}]{Stappers+1996}
{Stappers} B.~W.,  et~al., 1996, \mn@doi [\apjl] {10.1086/310148}, \href
  {https://ui.adsabs.harvard.edu/abs/1996ApJ...465L.119S} {465, L119}

\bibitem[\protect\citeauthoryear{{Toledo-Padr{\'o}n}
  et~al.,}{{Toledo-Padr{\'o}n} et~al.}{2020}]{Toledo-Padron+2020}
{Toledo-Padr{\'o}n} B.,  et~al., 2020, \mn@doi [\aap]
  {10.1051/0004-6361/202038187}, \href
  {https://ui.adsabs.harvard.edu/abs/2020A&A...641A..92T} {641, A92}

\bibitem[\protect\citeauthoryear{{Vasiliev} \& {Baumgardt}}{{Vasiliev} \&
  {Baumgardt}}{2021}]{Vasiliev+Baumgardt2021}
{Vasiliev} E.,  {Baumgardt} H.,  2021, \mn@doi [\mnras]
  {10.1093/mnras/stab1475}, \href
  {https://ui.adsabs.harvard.edu/abs/2021MNRAS.505.5978V} {505, 5978}

\bibitem[\protect\citeauthoryear{{Verbunt} \& {Freire}}{{Verbunt} \&
  {Freire}}{2014}]{Verbunt+Freire2014}
{Verbunt} F.,  {Freire} P. C.~C.,  2014, \mn@doi [\aap]
  {10.1051/0004-6361/201321177}, \href
  {https://ui.adsabs.harvard.edu/abs/2014A&A...561A..11V} {561, A11}

\bibitem[\protect\citeauthoryear{{Verbunt} \& {Hut}}{{Verbunt} \&
  {Hut}}{1987}]{1987_Verbunt+Hut}
{Verbunt} F.,  {Hut} P.,  1987, in {Helfand} D.~J.,  {Huang} J.~H.,  eds,
  International Astronomical Union Symposia Vol. 125, The Origin and Evolution
  of Neutron Stars. p.~187

\bibitem[\protect\citeauthoryear{{Vleeschower} et~al.,}{{Vleeschower}
  et~al.}{2022}]{Vleeschower+2022}
{Vleeschower} L.,  et~al., 2022, \mn@doi [\mnras] {10.1093/mnras/stac921},
  \href {https://ui.adsabs.harvard.edu/abs/2022MNRAS.513.1386V} {513, 1386}

\bibitem[\protect\citeauthoryear{{Weltevrede}}{{Weltevrede}}{2016}]{Weltevrede2016}
{Weltevrede} P.,  2016, \mn@doi [\aap] {10.1051/0004-6361/201527950}, \href
  {https://ui.adsabs.harvard.edu/abs/2016A&A...590A.109W} {590, A109}

\bibitem[\protect\citeauthoryear{{Yao}, {Manchester}  \& {Wang}}{{Yao}
  et~al.}{2017}]{YM2016}
{Yao} J.~M.,  {Manchester} R.~N.,   {Wang} N.,  2017, \mn@doi [\apj]
  {10.3847/1538-4357/835/1/29}, \href
  {https://ui.adsabs.harvard.edu/abs/2017ApJ...835...29Y} {835, 29}

\bibitem[\protect\citeauthoryear{{Zhang} et~al.,}{{Zhang}
  et~al.}{2022}]{Zhang_2022}
{Zhang} L.,  et~al., 2022, \mn@doi [\apjl] {10.3847/2041-8213/ac81c3}, \href
  {https://ui.adsabs.harvard.edu/abs/2022ApJ...934L..21Z} {934, L21}

\bibitem[\protect\citeauthoryear{{de Oliveira-Costa}, {Tegmark}, {Gaensler},
  {Jonas}, {Landecker}  \& {Reich}}{{de Oliveira-Costa}
  et~al.}{2008}]{dOliveira-Costa+2008}
{de Oliveira-Costa} A.,  {Tegmark} M.,  {Gaensler} B.~M.,  {Jonas} J.,
  {Landecker} T.~L.,   {Reich} P.,  2008, \mn@doi [\mnras]
  {10.1111/j.1365-2966.2008.13376.x}, \href
  {https://ui.adsabs.harvard.edu/abs/2008MNRAS.388..247D} {388, 247}

\bibitem[\protect\citeauthoryear{{van Haaften}, {Nelemans}, {Voss}  \&
  {Jonker}}{{van Haaften} et~al.}{2012}]{vanHaaften+2012}
{van Haaften} L.~M.,  {Nelemans} G.,  {Voss} R.,   {Jonker} P.~G.,  2012,
  \mn@doi [\aap] {10.1051/0004-6361/201218798}, \href
  {https://ui.adsabs.harvard.edu/abs/2012A&A...541A..22V} {541, A22}

\bibitem[\protect\citeauthoryear{{van Straten} \& {Bailes}}{{van Straten} \&
  {Bailes}}{2011}]{vanStraten+2011}
{van Straten} W.,  {Bailes} M.,  2011, \mn@doi [\pasa] {10.1071/AS10021}, \href
  {https://ui.adsabs.harvard.edu/abs/2011PASA...28....1V} {28, 1}

\makeatother
\end{thebibliography}

\bsp	
\label{lastpage}
\end{document}